\documentclass[authoryear,preprint,12pt,3p]{elsarticle}

\usepackage{lineno,hyperref}
\modulolinenumbers[5]

\usepackage{graphics}
\usepackage{amssymb}
\biboptions{comma,round,sort}

\usepackage[utf8]{inputenc}
\usepackage{gensymb}
\usepackage[dvipsnames]{xcolor}
\usepackage{url}
\usepackage{float}
\usepackage{subcaption}
\usepackage[none]{hyphenat}

\journal{Icarus}

\begin{document}

\begin{frontmatter}

\title{Global seasonal variations of the near-surface relative humidity levels on present-day Mars}
%\tnotetext[label0]{Global seasonal variations of the near-surface relative humidity levels on present-day Mars}

\author[label1]{Bernadett P\'al\corref{cor1}}
\address[label1]{Research Centre for Astronomy and Earth Sciences, Konkoly Thege Miklos Astronomical Institute, Budapest, Hungary}
\address[label2]{LMD, Institut Pierre Simon Laplace Université Paris 6, BP 99, 4 place Jussieu, 75005 Paris, FRANCE}
\address[label3]{NASA Goddard Space Flight Center, Greenbelt, MD 20771, United States}

%\cortext[cor1]{I am corresponding author}
%\fntext[label3]{I also want to inform about\ldots}
%\fntext[label4]{Small city}

\ead{pal.bernadett@csfk.mta.hu}
%\ead[url]{author-one-homepage.com}

\author[label1]{\'Akos Kereszturi}
%\ead{author.two@mail.com}

\author[label2]{Fran\c{c}ois Forget}
%\ead{author.three@mail.com}

\author[label3]{Michael D. Smith}
%\ead{michael.d.smith@nasa.gov}

\begin{abstract}
We investigate the global seasonal variations of near-surface relative humidity and relevant attributes, like temperature and water vapor volume mixing ratio on Mars using calculations from modelled and measurement data. We focus on 2 am local time snapshots to eliminate daily effects related to differences in insolation, and to be able to compare calculations based on modelling data from the Laboratoire de M\'et\'eorologie Dynamique Mars General Circulation Model with the observations of Mars Global Surveyor Thermal Emission Spectrometer. We study the seasonal effects by examining four specific dates in the Martian year, the northern spring equinox, summer solstice, autumn equinox and winter solstice. We identify three specific zones, where the near-surface relative humidity levels are systematically higher than in their vicinity regardless of season. We find that these areas coincide with low thermal inertia features, which control surface temperatures on the planet, and are most likely covered with unconsolidated fine dust with grain sizes less than $\sim$ 40$\mu$m. By comparing the data of relative humidity, temperature and water vapor volume mixing ratio at two different heights (near-surface, $\sim$ 23 m above the surface), we demonstrate that the thermal inertia could play an important role in determining near-surface humidity levels. We also notice that during the night the water vapor levels drop at $\sim$ 4 m above the surface. This, together with the temperature and thermal inertia values, shows that water vapor likely condenses in the near-surface atmosphere and on the ground during the night at the three aforementioned regions. This condensation may be in the form of brines, wettening of the fine grains or deliquescence. This study specifies areas of interest on the surface of present day Mars for the proposed condensation, which may be examined by in-situ measurements in the future. 
\end{abstract}


\end{frontmatter}

 %\linenumbers

\section{Introduction}
\label{sec:intro}

The occurrence of liquid water is one of the most important components we are seeking during the search for habitable locations outside planet Earth. Mars has been of key importance in the search for extraterrestrial life for decades now \citep{gargaud2011}. During its geological history there were periods when the planet likely had lakes \citep{cabrol1999,fassett2008,ehlmann2016}, rivers or maybe even larger bodies of liquid water \citep{tokano2005}. A recent discovery of organic molecules found at Gale crater provided further arguments that Gale crater was potentially habitable about 3.5 billion years ago \citep{kate2018}. However liquid water cannot be sustained in large volumes at the surface of present day Mars mainly due to the thin atmosphere, low temperature and general dryness. Based on recent results, liquid water might emerge on a microscopic scale on the surface of hygroscopic minerals \citep{nikola2018}, when the atmosphere shows elevated humidity and the temperature is suitable for various hygroscopic salts \citep{pal2017}. The Rover Environmental Monitoring Station (REMS) of the Curiosity rover showed through meteorological observations, that the necessary atmospheric conditions are met \citep{torres2015}. The existence of liquid water or brines \citep{zorzano2009} is indicated in several other observations as well, that might support some movement on the Martian surface in the form of streaks emanating from Dark Dune Spots \citep{kereszturi2012, kereszturi2016}, Recurring Slope Lineae \citep{ojha2015, mcewen2011} and changing droplet-like features on the leg of the Phoenix lander \citep{renno2009}. The main goal of this study is to explore the global behaviour of near-surface relative humidity during different seasons and daily cycles at different locations to improve our general understanding where and when liquid water could be sustained. It is also aimed at contributing to future mission planning by showing locations with elevated relative humidity levels and possible related processes, e.g. deliquescence.\\

Saturation and supersaturation have to occur occasionally on Mars to produce the observed surface condensation of water ice and the aphelion cloud belt. Supersaturation has been observed in the Martian atmosphere \citep{maltagliati2011b} by SPICAM onboard Mars Express mission. Some former models suggested supersaturation might be rare on Mars \citep{richardson2002a} because of instantaneous formation of ice \citep{montmessin2004}, or predicted it to occur at higher altitudes \cite{moores2011}. The occurrence of nighttime saturation has been suggested earlier \citep{savijarvi1995,davies1979}, however previously most works focused on saturation at a substantial elevation above the surface, causing cloud condensation \citep{gurwell2000} and fog condensation, especially during the night \citep{davies1979, moores2011}. Based on microwave temperature and water profiling at low- to mid-latitudes, saturation occurs below 10 km during northern spring/summer \citep{clancy1996}. The annual migration of saturation height has been confirmed by OMEGA data, showing 5-15 km height during aphelion, while it stretches up to 55 km during perihelion \citep{maltagliati2011}. Surface saturation produced condensation might happen at low latitudes because of the very cold nights on Mars \citep{carrozzo2009}. The spatial and temporal pattern of saturation migrates by climate changes too, probably it moves toward the equator during the increase of orbital tilt \citep{forget2006, madeleine2009, jakosky1985}. In the light of the available literature, this is the first time global near surface relative humidity was analyzed in such detail. \\

\section{Methods}
\label{sec:methods}

We investigated the global variations of the relative humidity at the surface by two different approaches. First, we examined model calculations by the Laboratoire de M\'et\'eorologie Dynamique Mars General Circulation Model (LMDZ GCM), detailed in \citep{forget1999}, including a water cycle as described in \citep{navarro2014}. Second, we looked at the measurements of Mars Global Surveyor Thermal Emission Spectrometer (MGS TES) \citep{smith2006} and calculated relative humidity values from the available temperature \citep{conrath2000}, pressure and water vapor data \citep{smith2002}. \\

\begin{figure}[H]
    \centering
    \includegraphics[width=\textwidth]{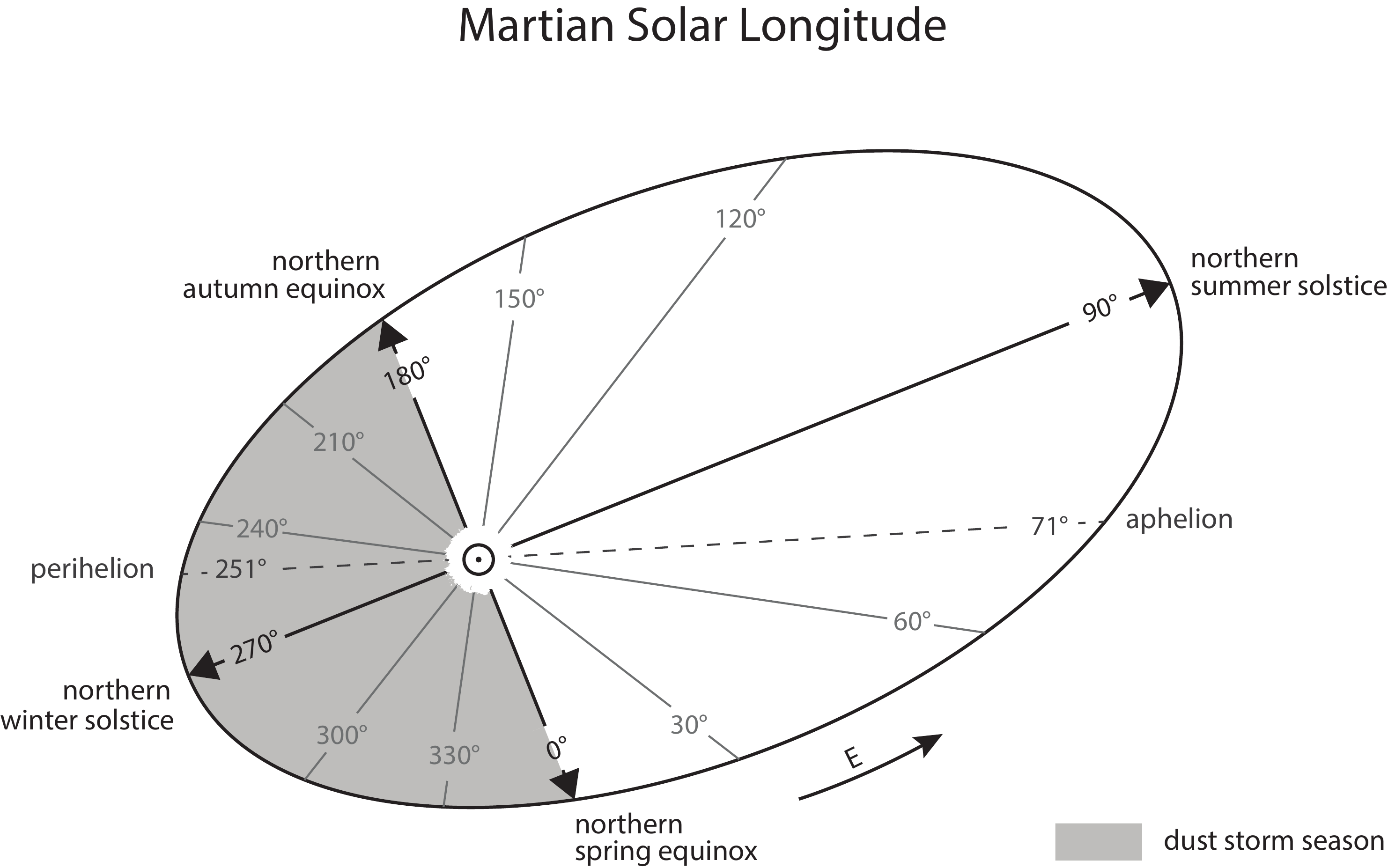}
    \caption{Martian solar longitude. A Martian year is 668.59 Martian days (sols) long, which is equal to 687 Earth days, because a sol is 24.623 hours long, $\sim$ 40 minutes longer than a day on Earth \citep{clancy2000}. To mark the time on Mars it is common to use the local time in Martian hours with a sol divided by 24 and the areocentric longitude of the Sun (Ls) as viewed from the centre of Mars. This is equivalent to ecliptic longitude.}
    \label{fig:solarlong}
\end{figure}

Our maps are represented on an equirectangular latitude-longitude grid. Each map is a snapshot of the planet, where the local time does not change with geographical longitude. These are composite maps with the same local time everywhere, e.g. if it is 2 am LT at 0$\degree$ E, it is also 2 am LT at 120$\degree$ W. The maps illustrating calculations from GCM modelled data are either at approximately 4 m above surface, or from approximately 23 m, always indicated in the titles. The 4 m height will be referred to in the following as values at the surface, because in calculating relative humidity values at the surface, we used surface temperature, surface pressure and the water vapor volume mixing ratio at 4 m height (the first atmospheric level in GCM), and presumed the water vapor to be well mixed between the surface and the first layer. \\

To get the \textbf{relative humidity} values, the saturation water vapor volume mixing ratio is needed. This can be calculated with the following equation used in the MCD model, which is based on the Tetens formula: 

\begin{equation}
    \label{eq:qsat}
    \mathrm{Q_{sat}} = \frac{100}{\mathrm{P}} \times 10^{2.07023 - 0.00320991 \mathrm{T} - \frac{2484.896}{\mathrm{T}} + 3.56654 \mathrm{log(T)}}
\end{equation}

\noindent where $\mathrm{P}$ is the surface pressure and $\mathrm{T}$ is the surface temperature. \\

After determining the necessary $\mathrm{Q_0}$ value, the relative humidity can be calculated as:

\begin{equation}
    \label{eq:rh}
    \mathrm{RH} = \frac{\mathrm{Q_0}}{\mathrm{Q_{sat}}}
\end{equation}

\subsection{GCM modelling data}
\label{subsec:gcmmet}

Our first approach in determining the global distribution of relative humidity was through modelling. For this, we used the data files created by the LMDZ GCM. The files contained simulated data for a whole Martian year, averaged from 3 simulated years, and consisted of a large amount of results, including, but not limited to surface temperature and pressure values, information about winds and water vapor volume mixing ratio at different altitude levels. These data files came in a NetCDF format which we handled with different C and C++ programs of our own and visualized with Panoply. \\

The water vapor volume mixing ratio (VMR) calculations are not reliable under $\sim$ 4 m due to the complexity of modelling near-surface interactions. For this reason in our study we used the $\sim$ 4 VMR data together with surface temperature and surface pressure values to estimate near-surface relative humidity by assuming the vapor to be well mixed between the surface and $\sim$ 4 m. Because we compare atmospheric vapor values with surface temperature, the relative humidity numerical values can reach unrealistically high outliers in some locations. Physically the supersaturation never goes very high, because the nighttime atmosphere is much warmer than the surface, and just near it the atmosphere is drier due to the interaction with the surface. In some way we show a ``potential supersaturation" assuming that the water vapour at $\sim$ 4 m reflects the value at the surface. While the specific numerical values might not be accurate in all locations, the distribution of elevated relative humidity is reliable, which is why we concentrate on the global scale in this study. \\

\subsection{TES data}
\label{subsec:tesmet}

The Mars Global Surveyor (MGS) was launched on November 7, 1996 and was successfully put into orbit around Mars on September 12, 1997. The mission ended on November 2, 2006, when the mission controllers lost contact with the spacecraft. The TES instrument measured and monitored the Martian surface and atmosphere throughout the mission and collected over 206 million infrared spectra of the thermal infrared energy (heat) emitted from Mars \citep{smith2004, smith2006}. This survey also provided the first detailed look at the surface composition of Mars. \\

To easily compare GCM results with the TES measurement data, we created a NetCDF format file from the TES data files as well. For the NetCDF format the data is required to be on an equally spaced grid, we modified the TES results by filling the absent measurement points with ``-1" values. The resulting NetCDF file has 207 Ls ($\Delta Ls = 5\degree$), 60 latitude ($\Delta lat = 3\degree$) and 48 longitude ($\Delta lon = 7.5 \degree$) grid points, and thus 596160 data points. The data was acquired between July, 1998 and January, 2006 from the 24th martian year to the 27th. The estimated uncertainty for water vapor abundance is $\pm$ 3 precip-microns or 10\%, whichever is greater. For the temperature data the uncertainty is approximately $\pm$ 2-4 K. \\

For the relative humidity values, we first needed to determine the water vapor mixing ratio. Details of the mixing ratio calculation can be seen in appendix \ref{app:tes}. The column mass of water vapor is: 

\begin{equation}
    \mathrm{m_{H_2O}} = 0.001 \times \mathrm{TES_{\,H_2O} \: column \, (kg\,m^{-2})}
\end{equation}

\noindent with $\mathrm{TES_{\,H_2O} \: column}$  measured by TES in precipitable microns. If we assume that water vapor is confined below a pressure level $\mathrm{p_{top}\,(Pa)}$, the mean mass mixing ratio between $\mathrm{p_{surf}}$ and $\mathrm{p_{top}}$ is: 

\begin{equation}
    \mathrm{mmr} = \frac{\mathrm{m_{H_2O}} \times \mathrm{g}}{\mathrm{p_{surf}} - \mathrm{p_{top}}}
    \label{eq:tesmmr}
\end{equation}

\noindent with gravity denoted by g ($\mathrm{m\,s^{-2}}$). The corresponding volume mixing ratio is: 

\begin{equation}
    \mathrm{Q_0} = \mathrm{mmr} \times \frac{44}{18}
    \label{eq:tesq0}
\end{equation}

\noindent with 44 $\mathrm{g\,mol^{-1}}$ and 18 $\mathrm{g\,mol^{-1}}$ being the molar mass of Martian air and water, respectively. \\

\noindent In this estimation, we assumed that water vapor is well mixed, which means that the mixing ratio is constant from the surface ($\mathrm{p_{surf}}$) to a certain height in the atmosphere ($\mathrm{p_{top}}$). The most simple way is to choose $\mathrm{p_{top}}$ to be equal to $0$. This means that there is no ``upper limit", the water vapor extends through the atmosphere. A more realistic approach would be if we take the pressure level as the value of $\mathrm{p_{top}}$ where the water vapor condensates, also known as the \textbf{condensation height} \citep{smith2002}. Knowing the condensation levels as a function of solar longitude and latitude, and choosing $10 \, \, \mathrm{km}$ as the pressure scale height: 

\begin{equation}
    \label{eq:ptop}
    \mathrm{p_{top}} = \mathrm{p_{surf}} \times \mathrm{e^{-H / 10 \, \,km}}
\end{equation}

\noindent where $\mathrm{H}$ denotes the condensation height given in kilometers. After calculating $\mathrm{Q_0}$ this way, equation \ref{eq:rh} can be used to determine the relative humidity directly above the surface. \\

Reliable water vapor retrievals have only been performed on the TES daytime observations, so in our TES calculations we used TES water vapor data from approximately 2 pm together with approximately 2 am TES surface temperature and pressure data. To estimate the validity of substituting daytime vapor measurements in nighttime calculations, we computed relative humidities for 3 Martian years using nighttime TES surface temperature and pressure data, and we took water vapor volume mixing ratio modelled first for 2 am and then for 2 pm from GCM. For this, we interpolated the GCM data to perfectly fit the solar longitude, latitude and longitude grid of TES observational data. Afterwards we used a simple statistical test to check the differences between the results: 

\begin{equation}
    \label{eq:valid}
    \mathrm{z} = \frac{\left| \mathrm{RH_{n} - RH_{d}} \right|}{\sqrt{\mathrm{RH_n + RH_d}}}
\end{equation}

\noindent where $\mathrm{RH_n}$ and $\mathrm{RH_d}$ are the relative humidities calculated with nighttime TES measurements together with 2 am GCM VMR and 2 pm VMR respectively. Then, we computed yearly average difference by summing up the differences for each grid point and dividing it by the number of data. Comparing these results to the annual averages of relative humidities, the variations are quite considerable. The RH values derived from using daytime GCM vapor data are consistently higher than the nighttime calculations, however, the overall global distribution is in good agreement. We concluded that while we possibly overestimate the scale of saturation (see Figure \ref{fig:valid}) by using 2 pm TES vapor data, our approach is valid to demonstrate the global distribution of elevated relative humidity. Thus in our study, the TES relative humidity maps are derived from $\sim$ 2 am surface temperature and surface pressure and $\sim$ 2 pm water vapor measurements. \\


\section{Results}
\label{sec:results}

In the following, we present the resulting figures showing relative humidity, temperature, surface temperature and water vapor volume mixing ratio for the entire planet. In the first subsection (\ref{subsec:gcmres}), we show the relative humidity values at the surface derived from the GCM model calculations by presenting 2 am LT global maps. This local time was selected, because the TES measured the surface temperatures at roughly 2 am and modelling for the same local times allows us to see global trends and variations coming from two different methods, thus giving us the full picture in greater detail. In these figures, we can examine the seasonal changes excluding daily variations. In subsection \ref{subsec:tesres} we demonstrate the relative humidity values calculated from nighttime TES measurement data, roughly 2 am everywhere for the four representative seasons. During all seasons at 2 am LT on the GCM model results (section \ref{subsec:gcmres}) three distinct areas appear as oversaturated regions, the first zone being encircled by Amazonis (Am), Alba Patera (AP) and Tharsis (T); the second around Arabia Terra (AT); and the third is the region around Elysium Mons (EM). These are of course only two zones in reality, but due to their appearance on the maps we refer them as three for easier identification.  We refer to these regions in short as the ``humid zones" from now on. Afterwards we review the validity of our TES calculations in \ref{subsec:validation}, then present the possible connection the surface thermal inertia in \ref{subsec:resti}. In \ref{subsec:tsandvmr} we examine the global behaviour of surface temperature and water vapor mixing ratio, two key elements in calculating the relative humidity, and in the end we discuss the importance of surface properties in \ref{subsec:surface} by comparing surface results with 23 m ones. \\

\subsection{GCM results}
\label{subsec:gcmres}

\begin{figure}[H]
    \centering
    \includegraphics[width=\textwidth]{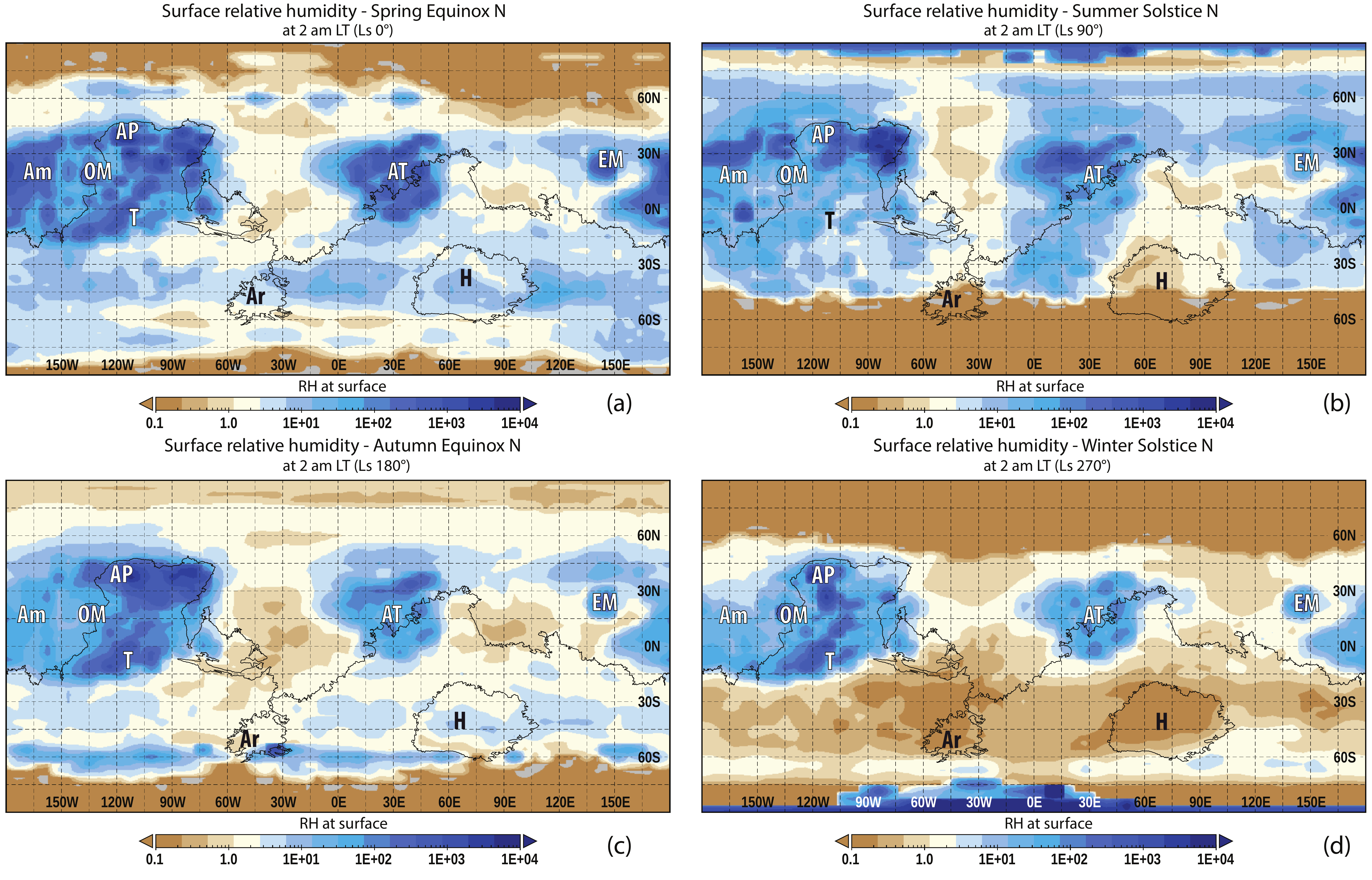}
    \caption{Seasonal dependence of the nighttime humidity distribution. We derived the relative humidity values directly above the surface from GCM model calculations at the time of northern spring equinox (a), northern summer solstice (b), northern autumn equinox (c) and northern winter solstice (d), 2 am local time at everywhere on the map. The three humid zones, defined in section \ref{sec:results} are visible in all seasons. They are the least defined during the time of northern summer solstice. During this time the regions south from Tharsis (T), Amazonis (Am) and Arabia Terra (AT) stretch further down south, with similar RH values. The three zones are visible the most clearly during northern winter, at which time their RH values are 2-3 magnitudes of order higher, than the rest of the planet surface.}
    \label{fig:gcm_rh_2am}
\end{figure}

A certain seasonal change exists in the distribution of relative humidity across the planet, as it is known that the Northern polar cap fills the atmosphere with vapor during the northern summer \citep{haberle1990}. The insolation-driven temperatures are related to the varying solar distance, and also differ according to the consequences of the tilt of the rotational axis. For this, we show four representative seasonal phases throughout the Martian year, the northern spring equinox (Ls 0$\degree$), summer solstice (Ls 90$\degree$), autumn equinox (Ls 180$\degree$) and winter solstice (Ls 270$\degree$). First, we look at the close surface relative humidity values at 2 am local time everywhere on the planet for these different seasons (Figure \ref{fig:gcm_rh_2am}). \\

At the northern spring figure (a) the three humid zones defined above are rather apparent in showing generally higher RH values than the rest of the planet. There is also a band, most clearly visible during northern spring and northern autumn between 30$\degree$ S and 55-60$\degree$ S, what seems more wet than its vicinity. The polar regions generally seem drier than anywhere else during the northern spring equinox, as water vapor in the atmosphere mainly starts to increase late spring (L$_s$= 45-90$\degree$) in the polar regions as a result of the sublimation of the seasonal ice deposits \citep{steele2014}. Comparing the northern spring to autumn, it is overall more humid during spring, excluding the three humid zones, which are more distinct during autumn than in spring. The wet band in the Southern Hemisphere appears here as well. Compared to the spring season, the northern polar region also has higher RH values during northern autumn. \\

The figure showing the 2 am snapshot at the time of the northern summer solstice (Figure \ref{fig:gcm_rh_2am}b) is quite different from the previous two seasons discussed, spring and autumn. While the three humid zones are there, they are not as well defined as in the other three seasons, and they are more extended. The distinct ``crescent moon" like shape east from EM, and EM is still distinguishable, but the regions around AT and T - Am - AP blend more into their environment. The overall location of the three zones are still the same, but the latter two humid zones stretch more to the north and south than they do in the other seasons. The southern polar region is dry, the wet band between 30$\degree$ S and 55-60$\degree$ S is missing here. During the northern winter on the other hand (Figure \ref{fig:gcm_rh_2am}d) the three zones are the most isolated in comparison with the other three seasons, the contrast between the RH values and the environment of the three humid zones appear here the strongest. The apparent ``borders" of these regions are well defined with OM and AP being the most humid. The southern pole seems rather wet, which could be the result of the sublimation of the ice during southern summer. Comparing the northern winter and northern summer it seems that the surface of Mars is generally higher in RH during the northern summer, in agreement with the expectations, as the atmospheric vapor is mainly released by the exposed water ice from the northern permanent polar cap. \\

\begin{figure}[H]
    \centering
    \begin{subfigure}[b]{0.49\textwidth}
    \includegraphics[width=\textwidth]{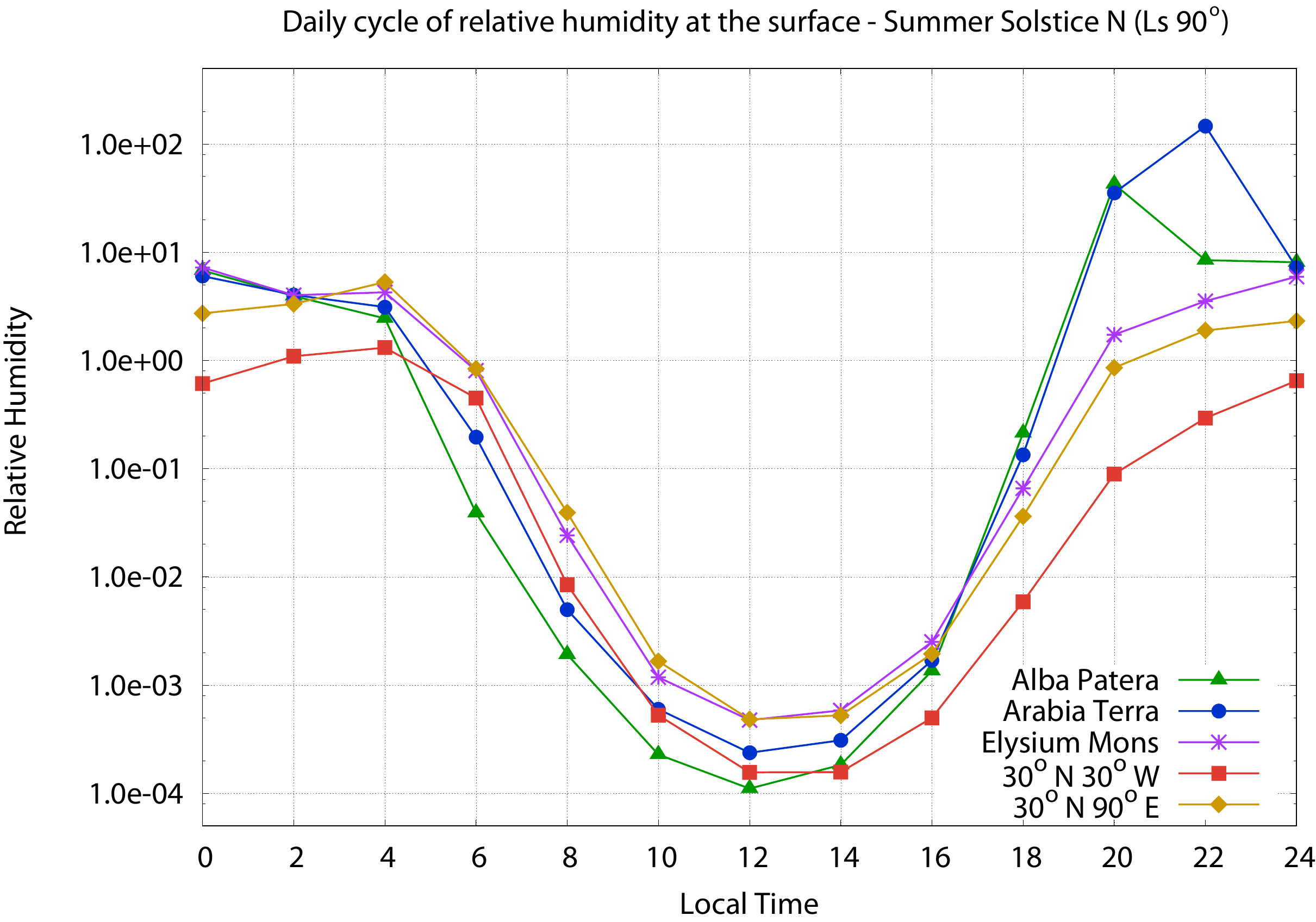}
    \end{subfigure}
    \begin{subfigure}[b]{0.49\textwidth}
    \includegraphics[width=\textwidth]{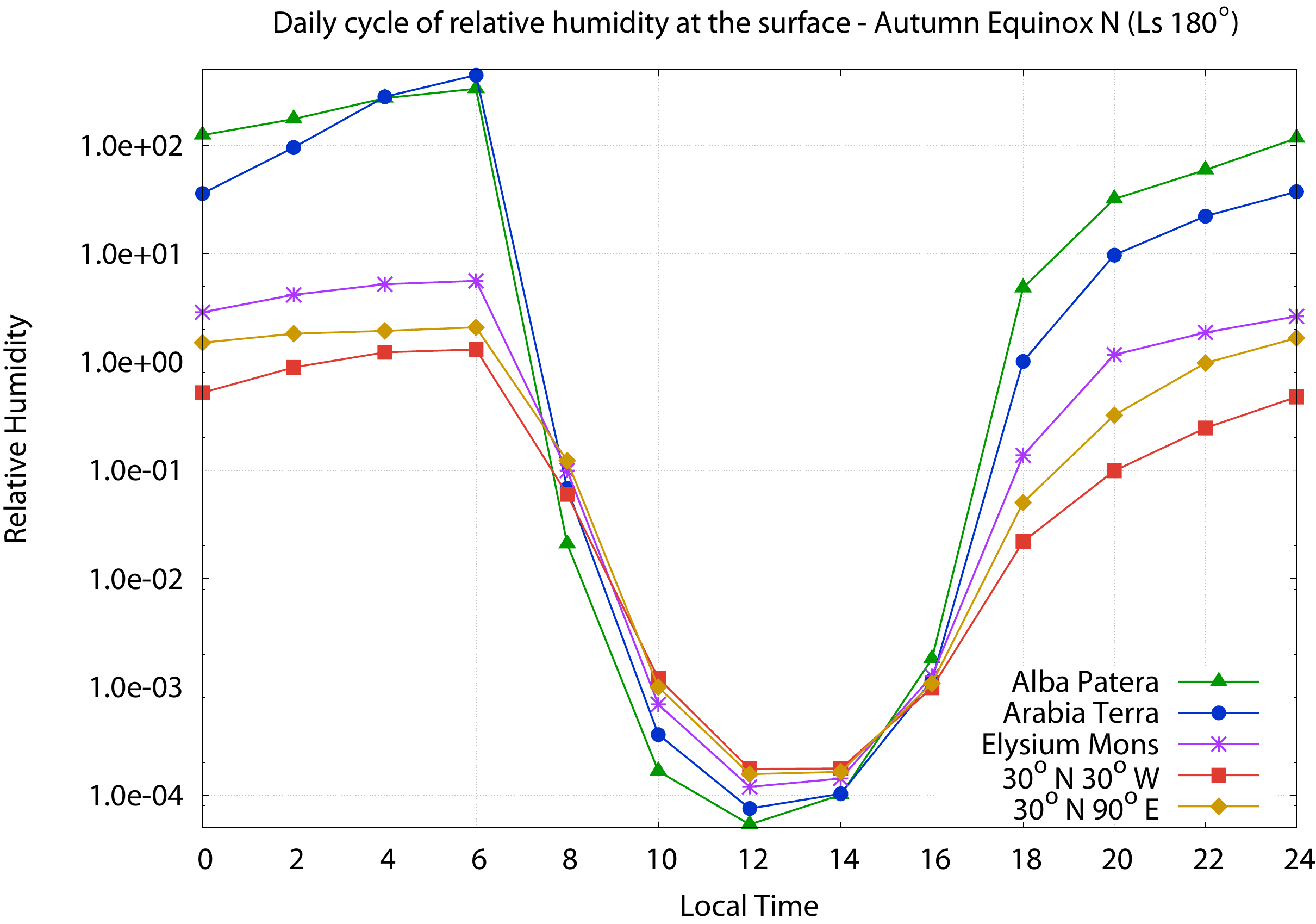}
    \end{subfigure}
    \caption{Daily cycles of near-surface humidity at the time of northern summer solstice and northern autumn equinox at Sol 195 (northern summer solstice - Ls 90) and at Sol 373 (northern autumn aquinox - Ls 180) calculated from GCM data. The five locations examined are Alba Patera (30$\degree$ N 120$\degree$ W), Arabia Terra (30$\degree$ N 30$\degree$ E), Elysium Mons (30$\degree$ N 150$\degree$ E) from the three previously defined humid zones, and two reference locations outside these humid zones, one located at 30$\degree$ N 30$\degree$ W and the other at 30$\degree$ N 90$\degree$ E.}
    \label{fig:rhs_SA}
\end{figure}

While we focus on snapshots mostly from 2 am in this study, it is interesting to take a look at the daily relative humidity trends, and the differences in them between the ``humid zones" and other regions. In Figure \ref{fig:rhs_SA} we can see daily relative humidity curves from five representative locations during northern summer solstice (a), and northern autumn equinox (b). The locations were picked out at the same latitude, in order to minimize variations due to differences in insolation. Three of the five sites are parts of the three humid zones defined in section \ref{sec:results}, AP (30$\degree$ N 120$\degree$ W), AT (30$\degree$ N 30$\degree$ E) and EM (30$\degree$ N 150$\degree$ E). We selected two reference locations, where the relative humidity levels show no significant elevation, at 30$\degree$ N 30$\degree$ W and at 30$\degree$ N 90$\degree$ E. We see, that elevated humidity levels with low temperatures during the night, and daytime dryness with higher temperatures are characteristic for all sites and seasonal periods. Comparing the northern summer solstice (Figure \ref{fig:rhs_SA}a) with the northern autumn equinox (b), we can see that the RH levels are more elevated during the night in autumn at AP and AT. The descending branches of the curves are steeper during autumn, the RH levels generally seem to start decreasing around 4 am during summer, but they stay elevated until 6 am in autumn, as the hours exposed to insolation are longer during summer. A similar effect is barely visible in the ascending branch, as the RH levels typically seem to start increasing around 4 pm during both seasons. AP and AT show higher levels by 6 pm in autumn than they do during summer, but there is no significant difference at EM and at the reference locations. \\

During autumn at AP and AT we observe that the RH levels start to increase around 4 pm and keep rising until 6 am in the morning. However, during summer they start to decline after 10 pm, then strongly decrease after 4 am. During autumn at AP and AT there is another striking feature, that the RH levels are the highest during the night, and they are the lowest during the day as well - possibly caused by the larger daily temperature fluctuation due to the lower thermal inertia. The same feature is not observable during summer, the RH levels do not differ as much between AT, AP and the other two locations. At EM and the reference locations there is no significant difference between summer and autumn, aside from the RH levels declining faster during summer morning. Comparing the reference locations to the other three regions, we can observe that the RH levels at 30$\degree$ N 30$\degree$ W and 30$\degree$ N 90$\degree$ E are changing less under a day overall, being usually lower during the night, and higher during the day than at the ``humid areas". \\

\subsection{TES results}
\label{subsec:tesres}

\begin{figure}[H]
    \centering
    \includegraphics[width=\textwidth]{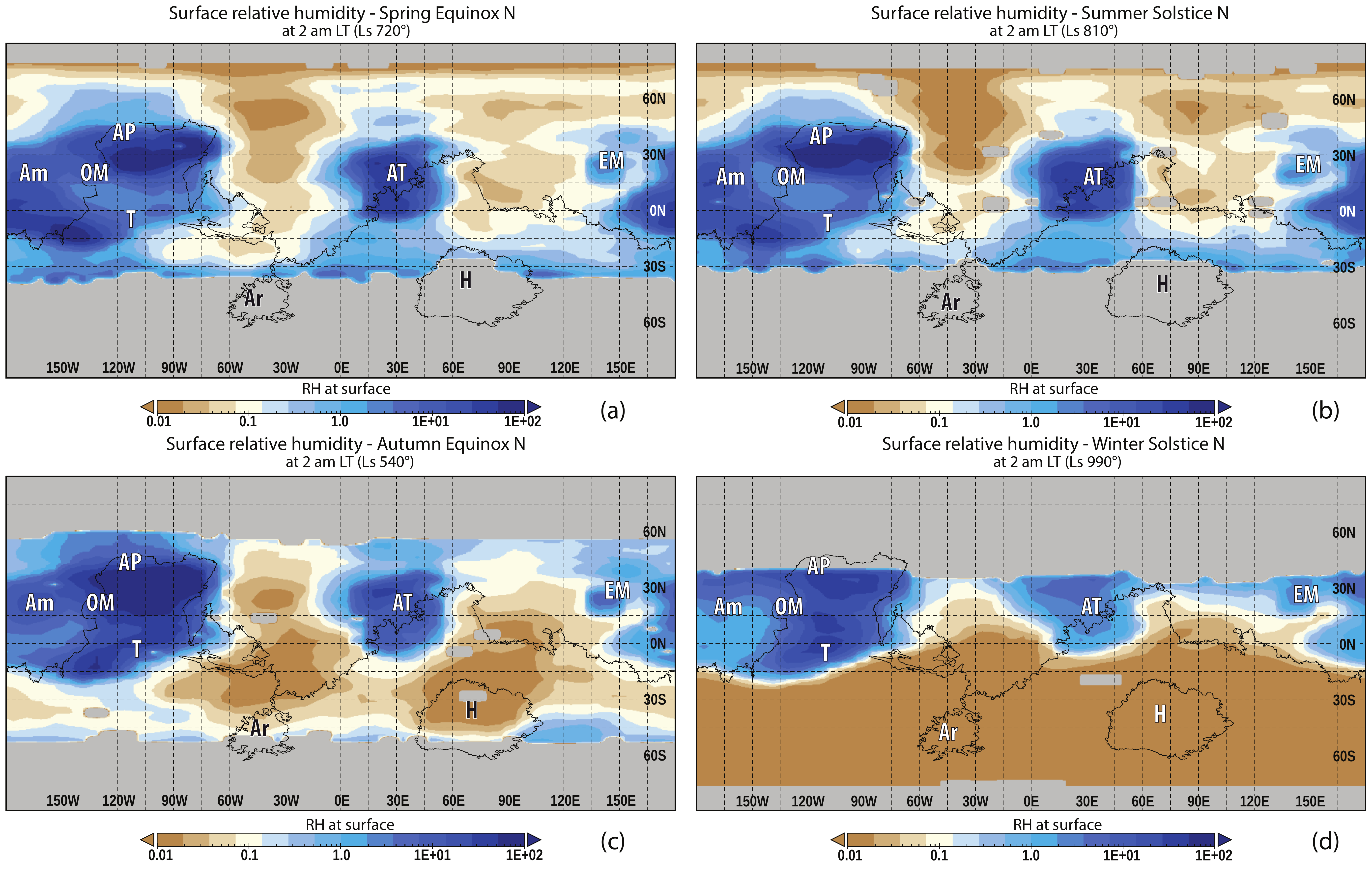}
    \caption{Seasonal dependence of the nighttime humidity distribution. We derived the relative humidity values from TES measurement data at the time of the northern spring equinox (a), northern summer solstice (b), northern autumn equinox (c) and northern winter solstice (d). The measured temperature and pressure data came from approximately 2 am local time, while the TES water vapor measurements were collected at approximately 2 pm. The three humid zones defined in section \ref{sec:results} are visible here as well. The regions showing the highest RH values are east from Alba Patera (AP) and around Arabia Terra (AT). These areas have about one order of magnitude higher RH values, than the other regions of the humid zones, and about 2-3 order of magnitude higher than the dry sectors. The gray areas in the figures correspond to places where relative humidity could not be calculated due to missing water vapor retrieval data.}
    \label{fig:tes_rh}
\end{figure}

The surface relative humidity values calculated from TES data are shown in Figure \ref{fig:tes_rh}, approximately 2 am local time everywhere for the four different seasonal times on Mars. The characteristics of the seasons are similar to what we can see on the data from GCM 2 am LT (Figure \ref{fig:gcm_rh_2am}). First, we look at the northern spring equinox in Figure \ref{fig:tes_rh}a, where the humid zones defined in section \ref{sec:results} are rather apparent. Out of these AT, the zone south from Am and the region east from AP show the highest RH values. There also seems to be a wet ``band" south from the equator (possibly one of the reasons behind this is the low temperature in that region), however since the condensation height could not be calculated due to missing water vapor retrieval data south from 55$\degree$S, the extent of the elevated RH band cannot be determined. Comparing northern spring to northern summer (Figure \ref{fig:tes_rh}b), we can see that the distribution of humidity levels is quite similar, the highest RH values show up at approximately the same regions, with the three zones again well defined. Please note that there are relatively dry areas between the wet regions at the same latitudinal band, for example between AP and AT locations. This demonstrates that the insolation or distance from the sublimating northern permanent cap is not the dominant influence producing elevated humidity values during the night. There is also indication for the wet ``band", but because of the missing data in the Southern Hemisphere, this is only a hint. \\

The northern autumn (Figure \ref{fig:tes_rh}c) and winter (Figure \ref{fig:tes_rh}d) figures are again similar to each other. The humid zones appear well defined in both. During autumn the Southern Hemisphere seems still quite humid, but during northern winter it seems almost completely dry. This latter observation might be related to both that the southern summer is warmer, and that large part of the the southern CO$_2$ cap is permanent, so the water ice exposure is limited during the local summer, resulting in an overall smaller contribution to the water vapor levels. In general, we can see that these are in good agreement with the GCM results shown in the previous sections, especially related to the location of the wet areas. \\

\subsection{Validation of TES calculations}
\label{subsec:validation}

There are no reliable water vapor measurements for 2 am, which is why we calculated near-surface relative humidity by using nighttime surface temperature and pressure data together with 2 pm vapor data. For this, we used TES water vapor measurements from 2 pm to calculate the water vapor volume mixing ratio ($Q_0$) according to Equation \ref{eq:tesq0}, and then determined relative humidity by diving $Q_0$ with $Q_{sat}$ (as seen in Equation \ref{eq:qsat}) calculated from 2 am pressure and surface temperature TES measurement data. The idea behind substituting 2 pm water vapor data as an approximation was, that surface temperature varies much more between day and night than water vapor varies between day and night. More precisely, the change in computed humidity using the observed daytime versus nighttime surface temperature is much greater than the change in computed humidity using the expected daytime versus nighttime water vapor abundance. We realize that water vapor is not the same during the day and night, this is an approximation, however at the present there is no good dataset giving the diurnal variation of water vapor for all seasons over a global scale. Therefore, we have chosen the simplest approximation: that there is no diurnal variation. A diurnal variation of 10-20\%, as in \citep{pankine2015} would not change any of the main findings of this work, including the location and seasonality of the ``humid zones". If we look at GCM derived column water vapor, the change from day to night is indeed small (less than 10\%). To check the validity of our approximation we followed a simple statistical approach detailed in section \ref{subsec:tesmet}. \\

\begin{figure}
    \centering
    \begin{subfigure}[b]{0.49\textwidth}
        \includegraphics[width=\columnwidth]{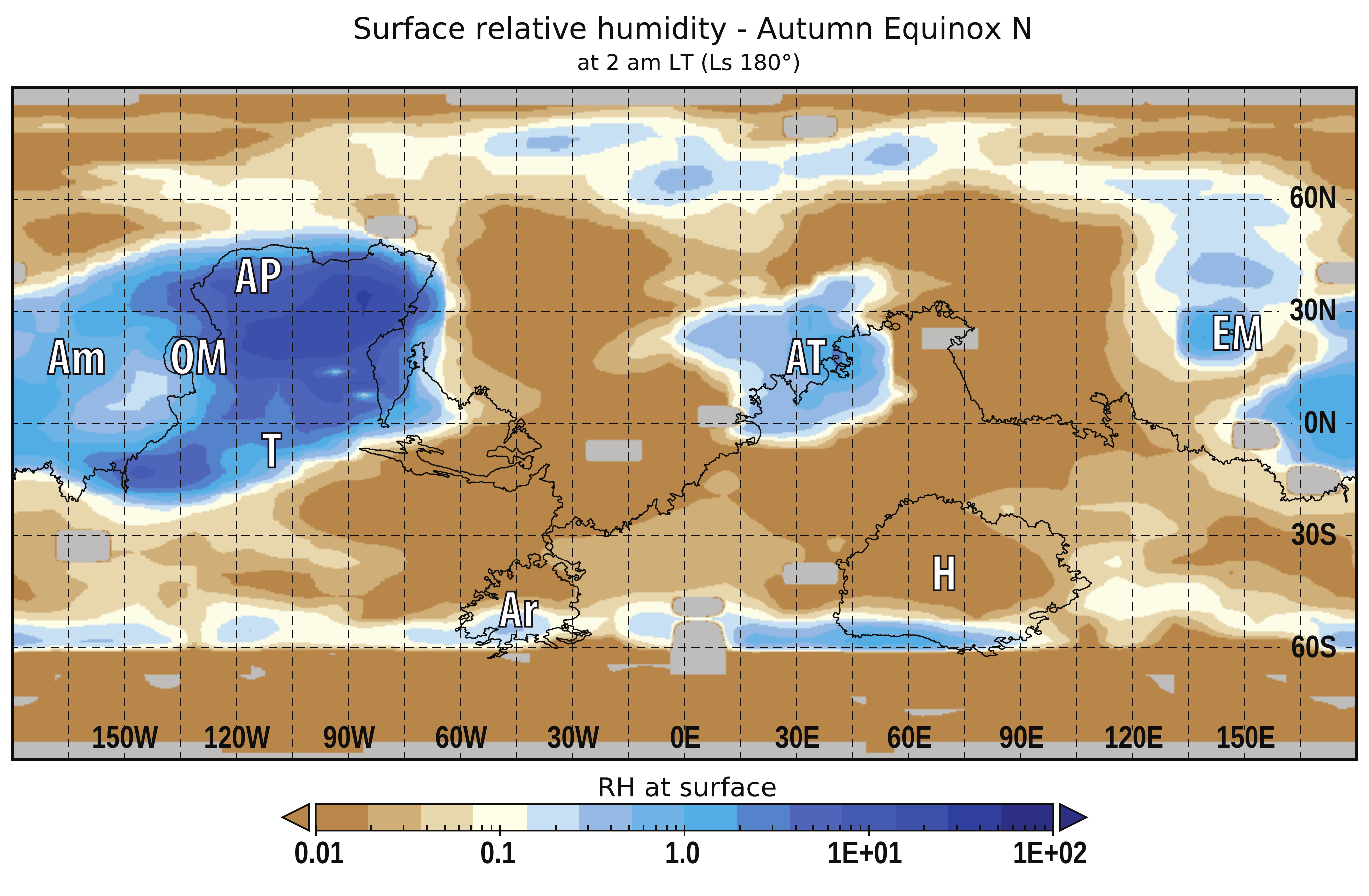}
    \end{subfigure}
    \hfill
     \begin{subfigure}[b]{0.49\textwidth}
        \includegraphics[width=\columnwidth]{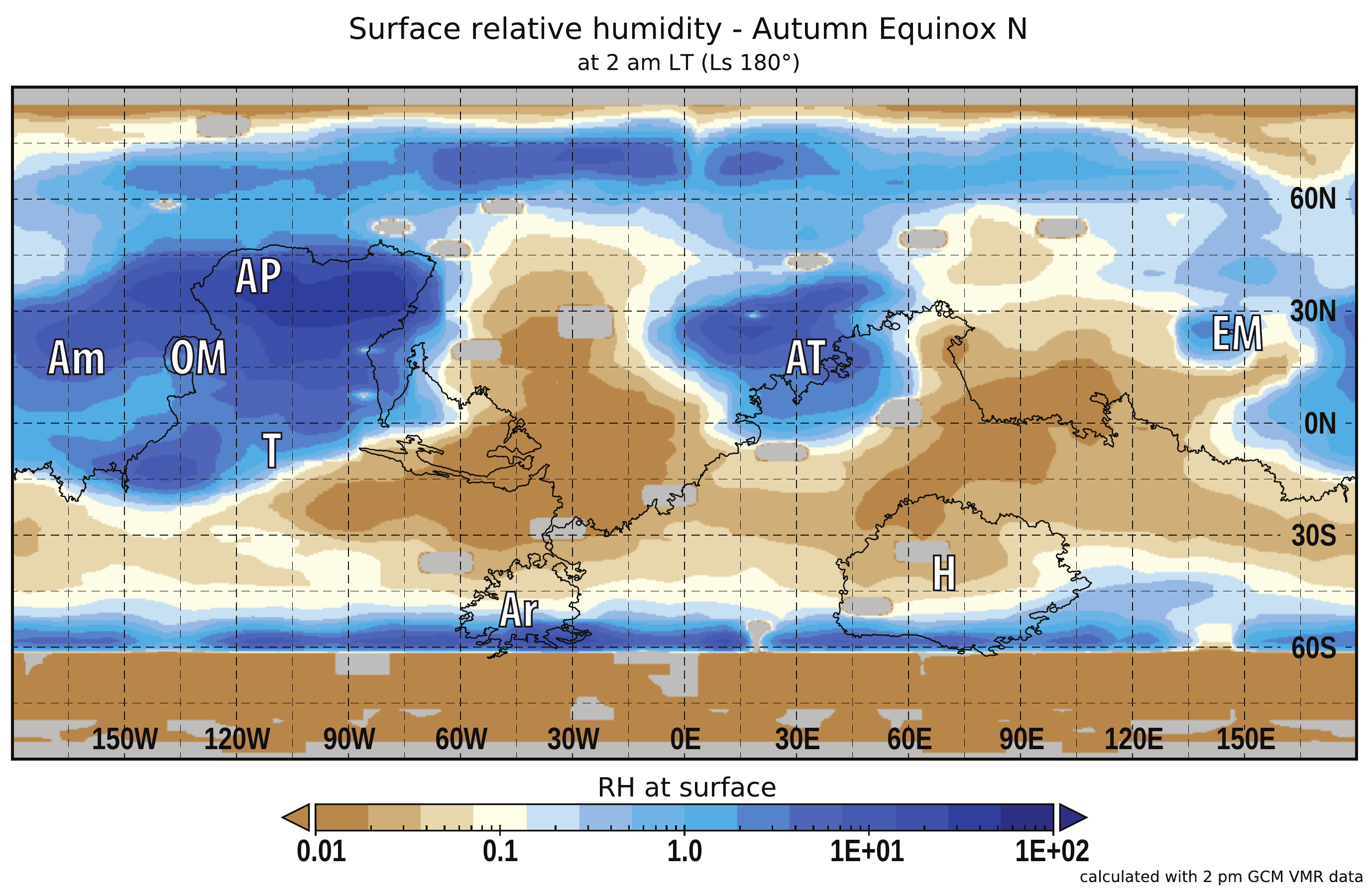}
    \end{subfigure}
    \caption{Differences of near-surface relative humidity if using nighttime or daytime water vapor data. In the left figure we can see the RH calculated with 2 am TES surface temperature and pressure measurements together with 2 am GCM VMR values. In the right we coupled the nighttime TES measurements with 2 pm GCM VMR data. It is evident that by using daytime water vapor we overestimate the saturation, but the global distribution of elevated RH regions remains the same.}
    \label{fig:valid}
\end{figure}

Figure \ref{fig:valid} shows the differences between near-surface relative humidity calculated with nighttime or daytime GCM vapor data to test the validity of our estimation. On the left we can see the GCM derived 2 am vapor data coupled with 2 am TES surface temperature and pressure values, while on the right we used 2 pm GCM vapor data with the nighttime TES measurements. Here, we present only the northern autumn season, but the outcome was the same for each season and TES measurement years. We most likely overestimate the scale of saturation by using daytime water vapor data, but the global distribution of regions showing elevated RH remains the same. The resulting maps are also consistent with the GCM results and TES results presented in the previous sections. Since the main goal of this paper is to investigate the global distribution of near-surface relative humidity and its variations between seasons and different geographical regions, we conclude that RH computed from TES this way is valid even with its known limitations and approximations. \\

\subsection{Relation to thermal inertia}
\label{subsec:resti}

\begin{figure}[H]
    \centering
%  \begin{subfigure}[b]{0.48\textwidth}
    \includegraphics[width=\textwidth]{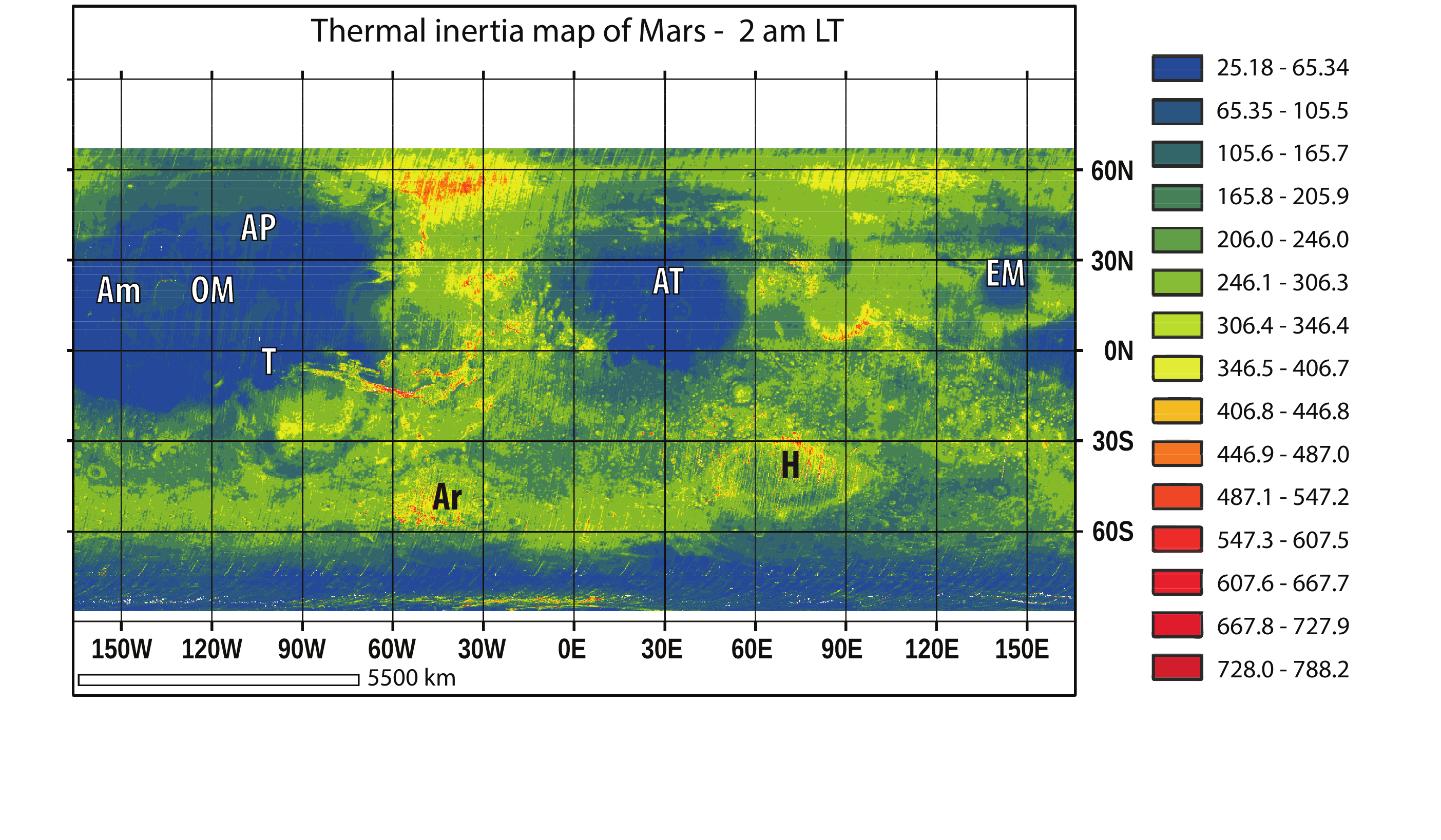}
    \caption{Thermal Inertia data at night measured by TES. The three zones defined in section \ref{subsec:gcmres} are visible here as low thermal inertia zones. The low thermal inertia area around Elysium Mons, as well as the slightly higher, but still low TI region north from there are also distinct and form the same specific shape resembling a crescent moon, as seen in the previous VMR, surface temperature and RH figures.}
    \label{fig:tesTI}
\end{figure}

Thermal inertia (TI) is one of the most important elements of diurnal temperature variations at the surface of Mars. It changes with particle size, rock abundance, degree of induration and exposure of bedrock within the top few centimeters of the subsurface. Thermal inertia provides a measure of the ability of the subsurface to store heat during the day and re-radiate it during the night \citep{putzig2005}. On the basis of TI values one can effectively estimate the grain sizes. Relatively higher thermal inertia values ($\sim$ 400 $Jm^{-2}K^{-1}s^{-1/2}$) indicate coarser material, while lower values ($\sim$ 250 $Jm^{-2}K^{-1}s^{-1/2}$) suggest fine grained particles like sand or dust ($\sim$ 20-100 $Jm^{-2}K^{-1}s^{-1/2}$). The temperature of a material with low thermal inertia values varies significantly during the day, while the change is not as drastic in case of materials with higher thermal inertia. In Fig. \ref{fig:tesTI} there are 3 distinct areas which have characteristically lower TI values than their surroundings. One is roughly between -10$\degree$E - 60$\degree$E, 5$\degree$S - 35$\degree$N, one is approximately between -60$\degree$E - 150$\degree$E, 25$\degree$S - 60$\degree$N, with a smaller region centered around 145$\degree$E, Elysium Mons. The third zone located below 60$\degree$S, and can be seen on all longitudes. \\

The first two regions (AP-OM-T and AT) in the Northern Hemisphere, and close to the equator shown in the thermal inertia map seem to be relatively in synch with the ``humid" zones observed in the TES and GCM 2 am maps. According to the research of \citep{putzig2005} these regions located mostly in the Northern Hemisphere are of low thermal inertia and high albedo, interpreted as surfaces dominated by unconsolidated fines (dust with grain sizes less than $\sim$ 40 $\mu m$). As for the band in the Southern Hemisphere, the surface is covered by material with relatively low thermal inertia and low albedo, comparable of the areas with thermal inertia around 300 $Jm^{-2}K^{-1}s^{-1/2}$. Normally such low thermal inertia values would suggest a surface of unconsolidated, fine-grained dust, such as the northern regions are believed to be, but because of the low albedo \citep{putzig2005} propose a different explanation: they conclude that a low-density mantle formed by desiccation of a previously ice-rich near-surface layer is the most likely explanation for the observed properties, a phenomena analogous to those which produce terrestrial low-density soils such as the desiccation of ice-cemented loess \citep{johnson2000}. \\

\subsection{Surface temperature and water vapor behaviour}
\label{subsec:tsandvmr}

\begin{figure}[H]
    \centering
    \includegraphics[width=\textwidth]{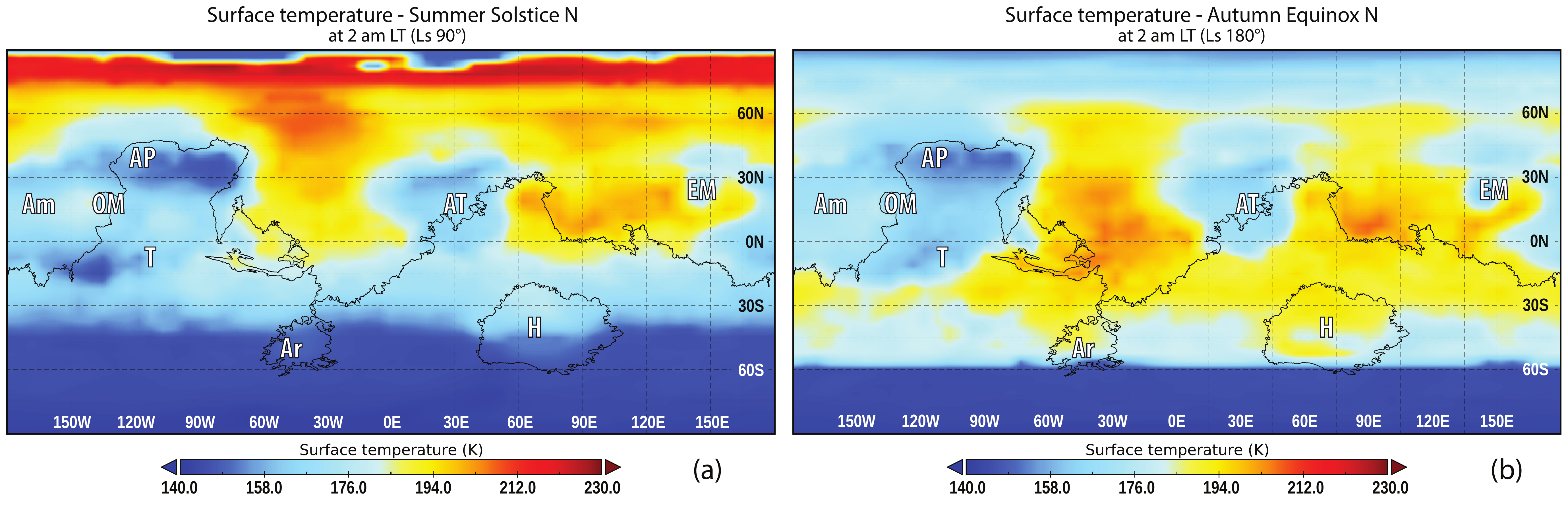}
    \caption{Seasonal dependence of the nighttime surface temperatures data from GCM at the time of the northern summer solstice (a) and northern autumn equinox (b), 2 am LT everywhere. The three ``humid" zones defined in section \ref{sec:results} are apparent here, these are approximately 30 K colder than their surroundings. The area east of Alba Patera (AP) seem the coldest part of these three zones, being an extra 10 K lower in temperature than the rest of the previously defined ``humid" regions.}
    \label{fig:gcm_ts}
\end{figure}

As the calculated relative humidity depends on temperature, pressure and water vapor volume mixing ratio (VMR), it is important to examine the characteristics of these parameters in understanding the global variations in RH. While the surface pressure values do change throughout the seasons, there is no global distribution variation similar to the surface temperature and water vapor volume mixing ratio changes, therefore in the following we will focus on the latter two parameters. First we look at the global characteristics of surface temperature during the night. Figure \ref{fig:gcm_ts} depicts the surface temperature values at 2 am local time everywhere on the planet, at the time of the northern summer solstice on the left (a), and northern autumn equinox on the right (b). In these figures it can be seen that three distinct areas in the Northern Hemisphere (Am-OM-AP, AT and EM) have approximately 30 degrees lower temperatures than their environment. These three zones correspond to the three humid zones defined in section \ref{sec:results}. During summer the area east of Alba Patera, and the region west from Tharsis seems to be the coldest of these three zones. \\

At the time of northern summer these zones are similar in temperature to the Southern Hemisphere, where it is currently winter. During northern winter the coldest regions seems to be around Olympus Mons and Alba Patera. The difference in insolation is also observable, as in the northern autumn the equatorial region is the warmest at night, however the three zones show little, only around 5-10 K variation in their temperature. This is still valid for the other two seasons not shown here, Ls 0$\degree$ and Ls 270$\degree$, thus we can conclude that these zones are consistently colder at 2 am, thus cool down during the night at any season throughout the Martian year.   \\

\begin{figure}[H]
    \centering
    \includegraphics[width=\textwidth]{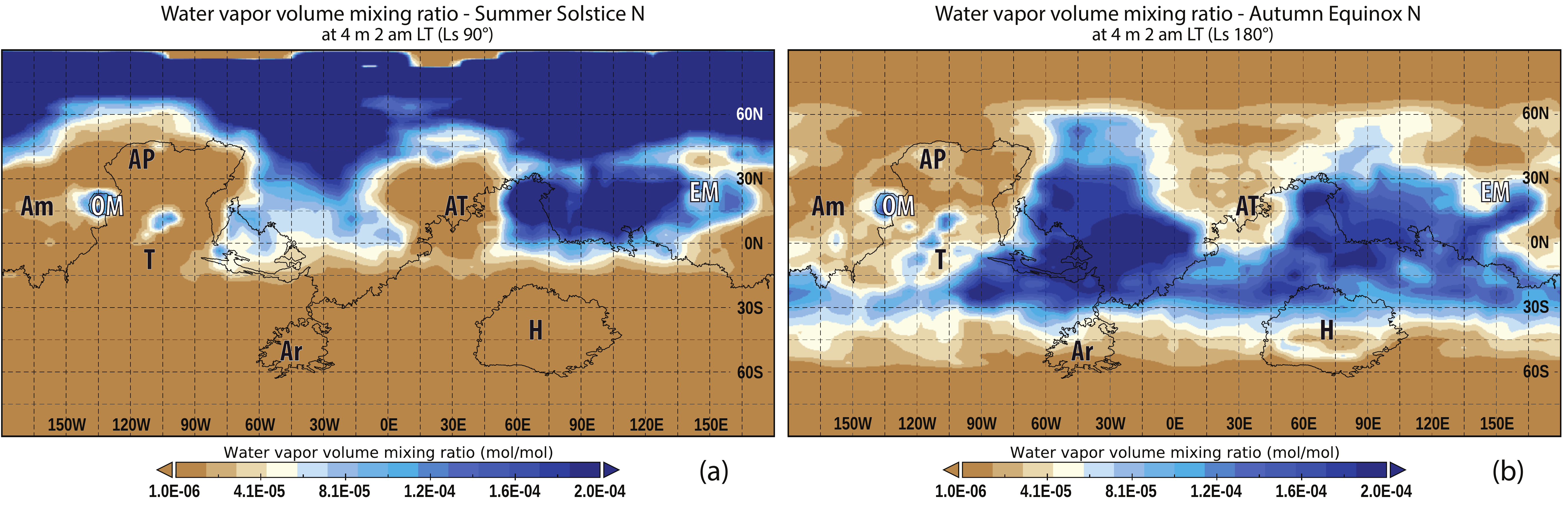}
    \caption{Seasonal dependence of water vapor volume mixing ratio data from GCM at the northern summer solstice (a) and northern autumn equinox (b) at approximately 4 m above surface, 2 am LT globally. The three humid zones defined in Section \ref{sec:results} appear here as regions, where the VMR is approximately 3 orders of magnitude lower than their surroundings. The exceptions are Olympus Mons (OM), Tharsis Montes (northwest from Tharsis denoted by T), and Elysium Mons (EM). These mountains show 1-2 orders of magnitude higher VMR values than the rest of the humid zones. We can also observe, that during northern summer the southern borders of these zones are not defined, with the exception of Arabia Terra (AT). During northern autumn the northern boundaries blend together with the northern polar region. This shows a trend of higher VMR values moving to the south as the seasons change from summer to eventually winter in the Northern Hemisphere \citep{smith2002,steele2014}.}
    \label{fig:gcm_vmr}
\end{figure}

Figure \ref{fig:gcm_vmr} illustrates the VMR snapshot from 2 am LT at approximately 4 m above the surface, with the northern summer solstice on the left (a) and the autumn equinox on the right (b). The three zones introduced in Section \ref{sec:results} are visible here as regions, where the VMR is lower than elsewhere in the Northern Hemisphere at the time, e.g. there is less water vapor in the near-surface atmospheric layers above these regions. The exceptions are Olympus Mons, Tharsis Montes and Elysium Mons, which show higher humidity than the rest of the ``humid zones". During the northern autumn shown in Figure \ref{fig:gcm_vmr}b, the three zones appear as rather vapor poor areas as well, though more humid, showing values of around 10$^{-5}$ mol/mol, while it is roughly an order of magnitude lower during summer. The regions around Olympus Mons, Tharsis Montes and Elysium Mons are again showing higher VMR values, with Tharsis Montes more defined here than during summer. This correlation with surface pressure (topography) have been observed previously also \citep{smith2002}, and it might still be true after the topography has been ``removed", because the water vapor is not completely well-mixed through the entire atmosphere. From the TES observations we find that a steep latitudinal gradient forms during aphelion season high in the north in water vapor abundance, while during the perihelion the water vapor is more uniformly spread out \citep{smith2002}. This phenomenon is observable here with the northern polar and high latitude regions showing higher VMR values than the equatorial regions at the time of the northern summer solstice (\ref{fig:gcm_vmr}a). The northern summer is characterized by a gradual southward transport of sublimed water from the northern polar cap to the equatorial regions \citep{steele2014} and by the time of the northern autumn equinox (\ref{fig:gcm_vmr}b) water vapor reaches peak values around the equator, corresponding to travel via Hadley cell circulation.\\

\begin{figure}[H]
    \centering
    \begin{subfigure}[b]{0.49\textwidth}
    \includegraphics[width=\columnwidth]{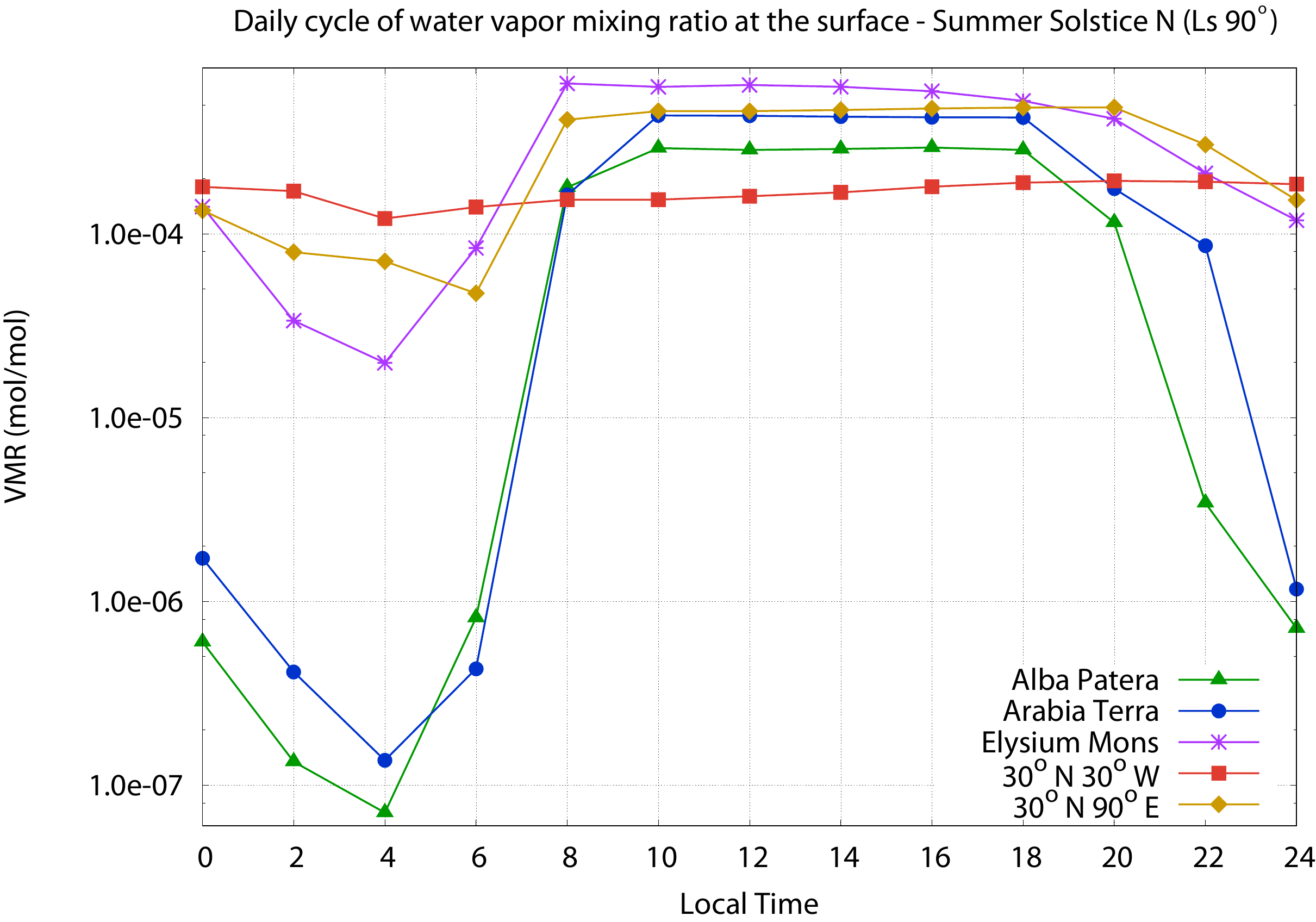}
    \end{subfigure}
    \begin{subfigure}[b]{0.49\textwidth}
    \includegraphics[width=\columnwidth]{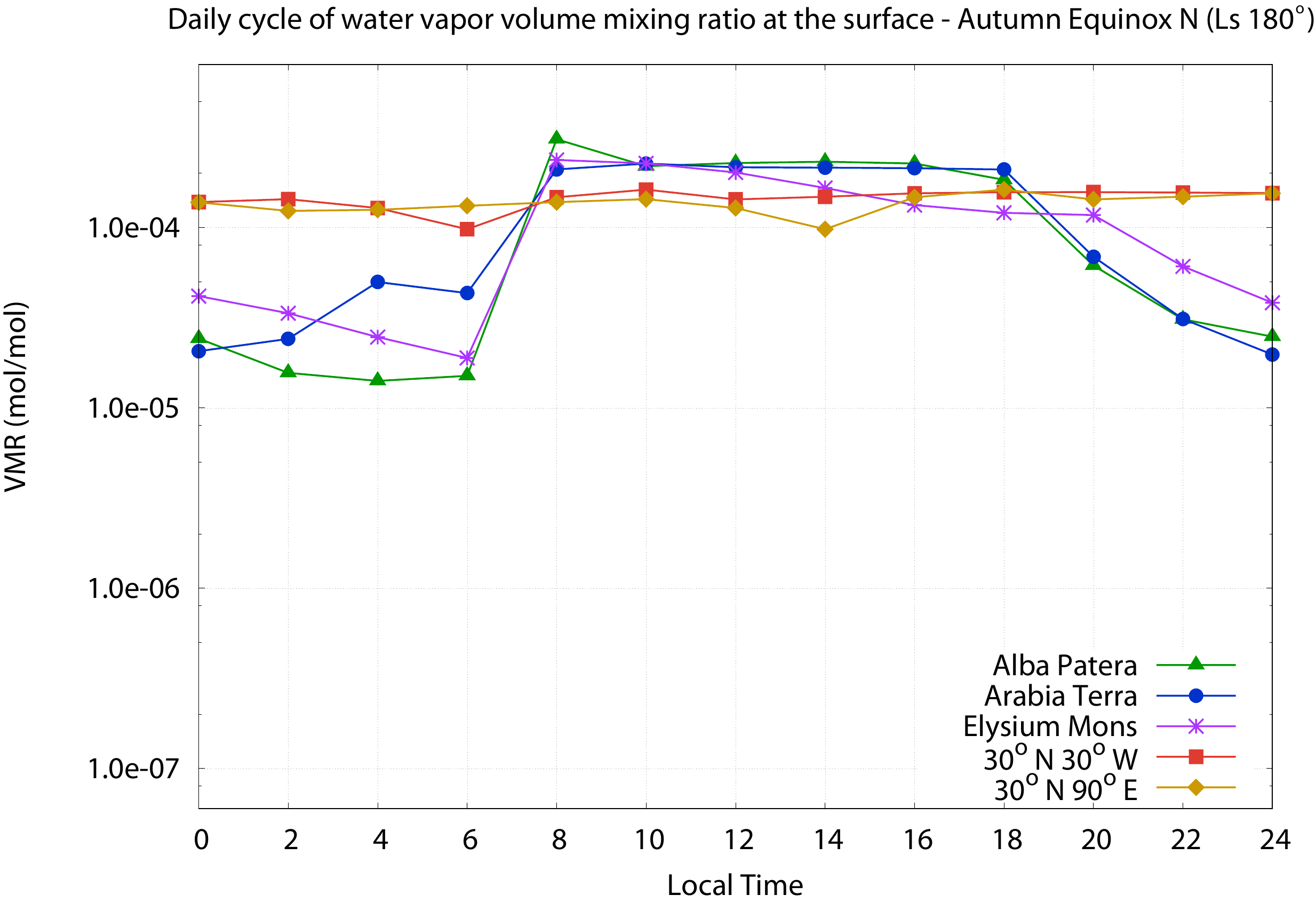}
    \end{subfigure}
    \caption{Daily curves of water vapor volume mixing ratio at the time of the northern summer solstice (left) and the northern autumn equinox (right). The locations are Alba Patera, Arabia Terra, Elysium Mons and two reference ``dry" locations, one at 30$\degree$ N 30$\degree$ W, and the other at 30$\degree$ N 90$\degree$ E. All selected analyzed points are located at latitude 30$\degree$ N to exclude different seasonal effects. The two reference ``dry" regions show smaller fluctuations throughout the day in both seasons.} 
    \label{fig:vmrd}
\end{figure}

\begin{figure}[H]
    \centering
    \includegraphics[width=\linewidth]{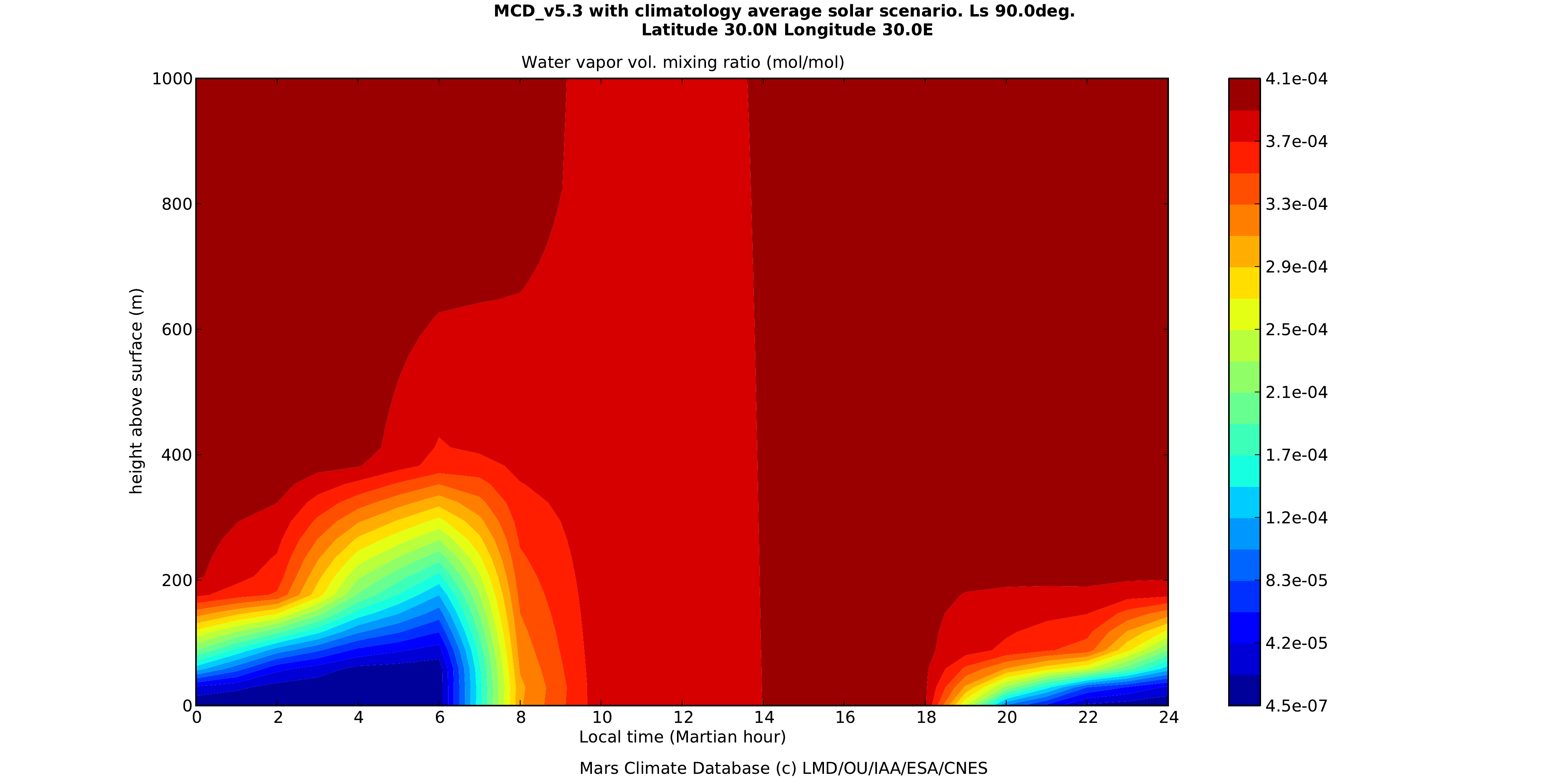}
    \caption{Daily variation of GCM derived water vapor volume mixing ratio with altitude at Alba Patera (30$\degree$N 30$\degree$E). A depletion is visible in the lower ranges of the atmosphere reaching the lowest values below 70 m. The water vapor volume mixing ratio is lowest between 8 pm and 6 am approximately.}
    \label{fig:depletion}
\end{figure}

To further examine the possible processes, we investigated the daily variations of water vapor volume mixing ratio above the surface at five representative locations at the same latitude to rule out seasonal effects. Three locations from the ``humid zones", 30$\degree$ N 30$\degree$ E from Arabia Terra, 30$\degree$ N 150$\degree$ E from Elysium Mons, 30$\degree$ N 120$\degree$ W from Alba Patera and two reference drier ares, one at 30$\degree$ N 30$\degree$ W and the other at 30$\degree$ N 90$\degree$ E. These regions cool down more quickly during the night than the ones with higher thermal inertia values. By examining the changes in water vapor volume mixing ratio at approximately 4 m above the surface, we can see in Figure \ref{fig:vmrd} that above the ``humid" areas, the VMR values drop during the night. If we look at the available water vapor with respect to the elevation (Figure \ref{fig:depletion}) it is clear that there is a depletion of water vapor near the surface at night (a couple of orders of magnitude difference). \\

To investigate whether the daily cycle of VMR shown in Figure \ref{fig:vmrd} is indeed a global behaviour and the depletion seen in Figure \ref{fig:depletion} is a global phenomenon outside 2 am as well, we included maps at two other representative local times. \\

\begin{figure}[H]
    \centering
    \begin{subfigure}[b]{0.49\textwidth}
    \includegraphics[width=\columnwidth]{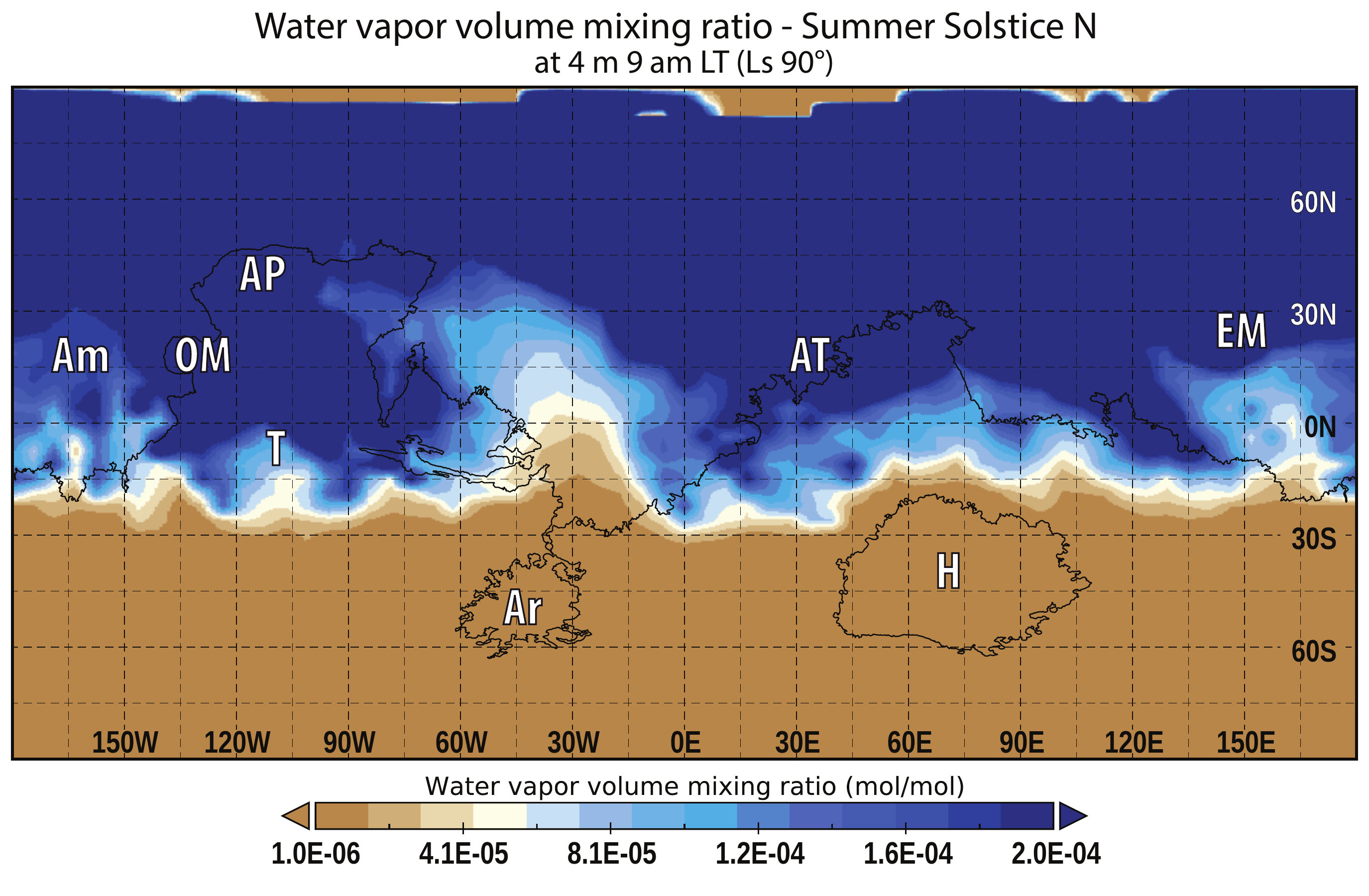}
    \end{subfigure}
    \begin{subfigure}[b]{0.49\textwidth}
    \includegraphics[width=\columnwidth]{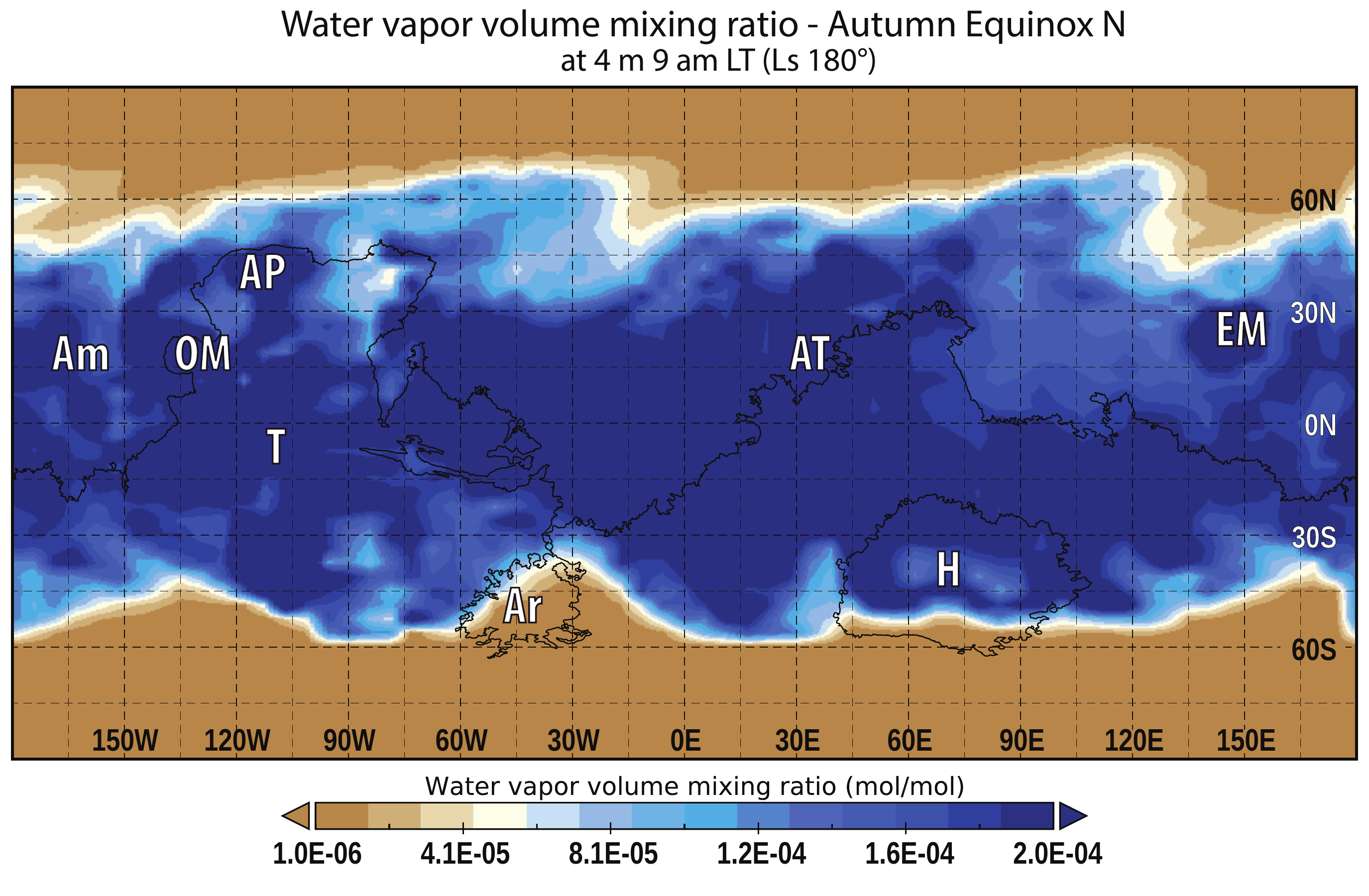}
    \end{subfigure}
    \caption{Global distribution of water vapor volume mixing ratio at 9 am LT derived from GCM at the time of the northern summer solstice (left) and northern autumn equinox (right). According to the daily curves, the VMR returns to the daytime levels by this time. In the map we can see, that indeed the water vapor above the surface is mostly continuous in distribution with values around 1.6$\times 10^{-4}$ mol/mol in both seasons presented here.} 
    \label{fig:vmr9am}
\end{figure}

\begin{figure}[H]
    \centering
    \begin{subfigure}[b]{0.49\textwidth}
    \includegraphics[width=\columnwidth]{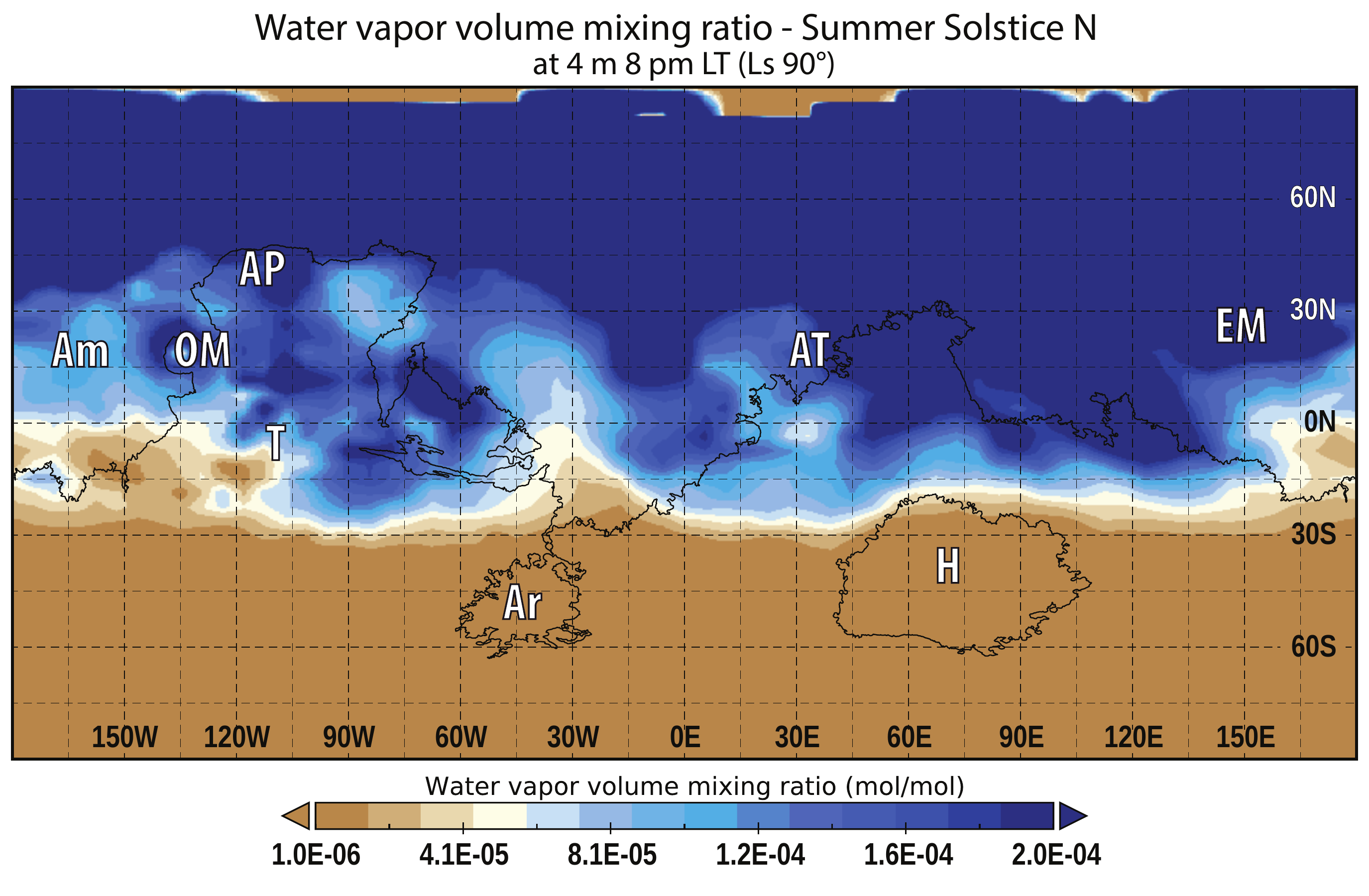}
    \end{subfigure}
    \begin{subfigure}[b]{0.49\textwidth}
    \includegraphics[width=\columnwidth]{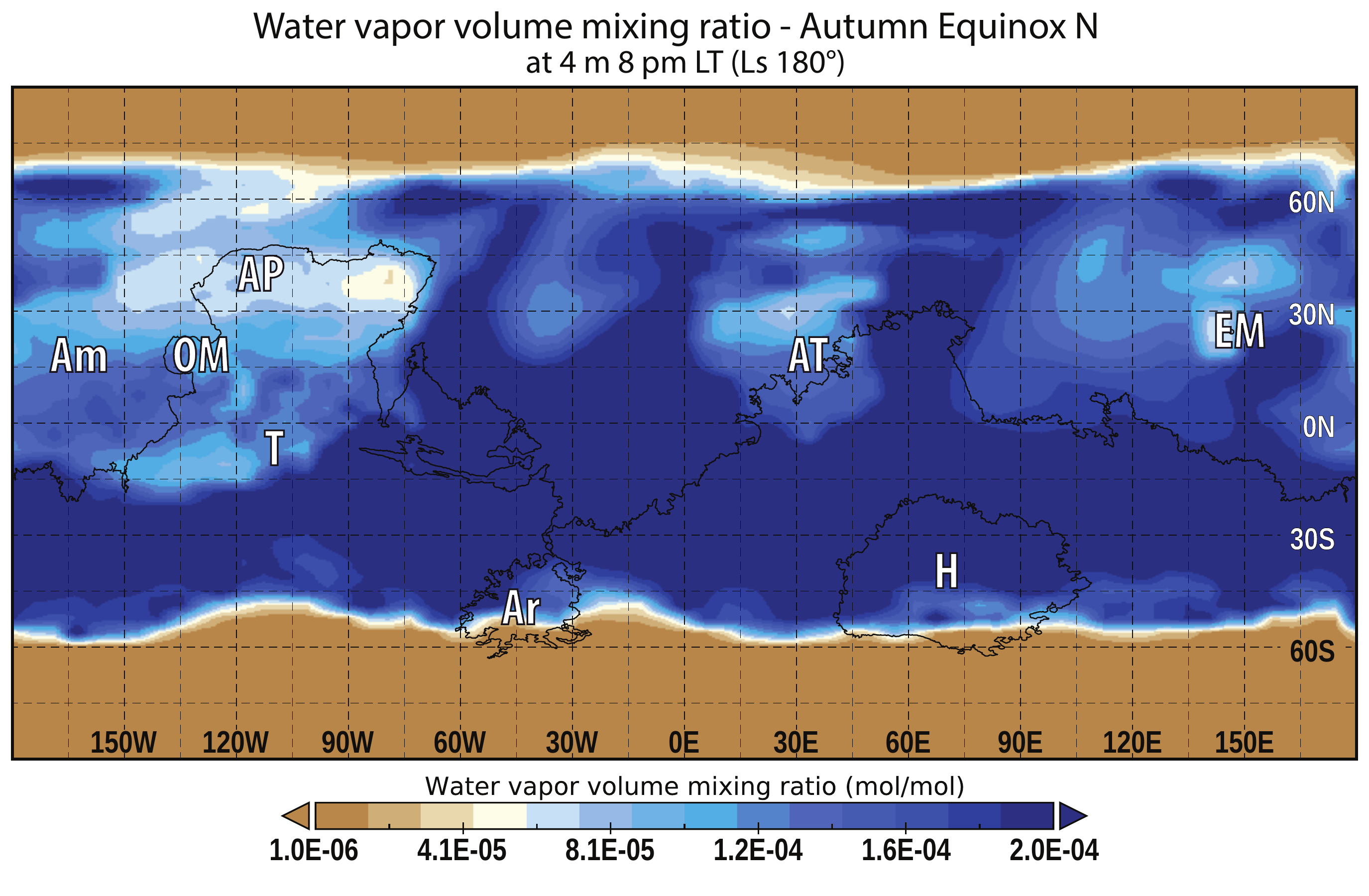}
    \end{subfigure}
    \caption{Global distribution of water vapor volume mixing ratio at 8 pm LT derived from GCM at the time of the northern summer solstice (left) and northern autumn equinox (right). The daily curves show, that at 8 pm the water vapor levels start to reduce, but not reach the minimum values yet. We can see that the vapor levels lower to around 1.0$\times 10^{-4}$ mol/mol above Am - AP - T, AT and in the vicinity of EM as well, while in other areas it mostly stays above 1.6$\times 10^{-4}$ mol/mol at the northern autumn equinox. The appearance of the ``humid zones" is less clear at the northern summer solstice due to overall higher daytime VMR in summer falling under 1.0$\times10^{-4}$ mol/mol approximately 1 hour later.}
    \label{fig:vmr8pm}
\end{figure}

Figures \ref{fig:vmr9am} and \ref{fig:vmr8pm} depict the changes of 4 m water vapor volume mixing ratio values in a global scale. The depletion of vapor during the night shown in Figure \ref{fig:depletion} is well observable not only in the daily curves (Figure \ref{fig:vmrd}), but looking at the whole near-surface of Mars as well. From the daily variations we have chosen 9 am LT and 8 pm LT to represent the global changes, since by 9 am the VMR levels rise to their daytime values, and by 8 pm they start to lower down at all locations investigated in Figure \ref{fig:vmrd}. At 9 am the water vapor at 4 m is mostly continuous in distribution with values reaching 1.6$\times10^{-4}$ mol/mol excluding the polar regions at the autumn equinox, and most of the southern hemisphere during the northern summer solstice. The vapor distribution does not change much during the daytime. As the night comes the vapor levels start to lower and Figure \ref{fig:vmr8pm} shows that VMR drops more quickly above the low TI ``humid zones". While the same difference in VMR is not that obvious at the time of the northern summer solstice due to the overall higher daytime VMR levels, we can see that compared to the 9 am figures, the vapor starts to condense from the lower atmosphere. Comparing all the water vapor figures at 4 m we can see, that VMR drops during the night to around 1.2$\times10^{-6}$ mol/mol by 6 am and it lowers more quickly above the ``humid zones". \\

\subsection{The importance of surface properties}
\label{subsec:surface}

\begin{figure}[H]
    \centering
    \includegraphics[width=\textwidth]{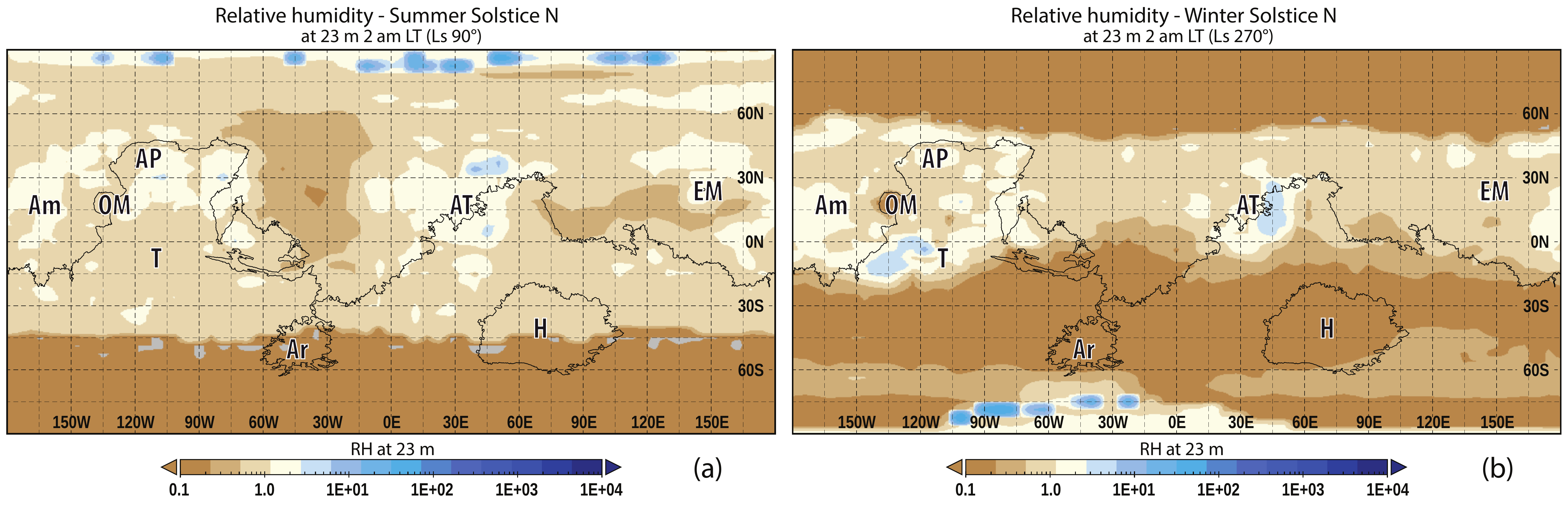}
    \caption{Seasonal dependence of nighttime humidity distribution at 23 m above the surface. The relative humidity values derived from GCM model calculations. Local time is 2 am globally at the time of northern summer solstice (a) and northern winter solstice (b). While there is a humid band visible with RH values around one order of magnitude higher than the dry areas during northern winter, the three humid zones defined in Section \ref{sec:results} do not appear here. We can also note the area between 60$\degree$ W - 15$\degree$ W with one order of magnitude lower RH values during northern summer. This is the same area, which was visible at the previous RH figures, but it disappears during the northern winter.}
    \label{fig:gcm_rh23m}
\end{figure}

\begin{figure}[H]
    \centering
    \includegraphics[width=\columnwidth]{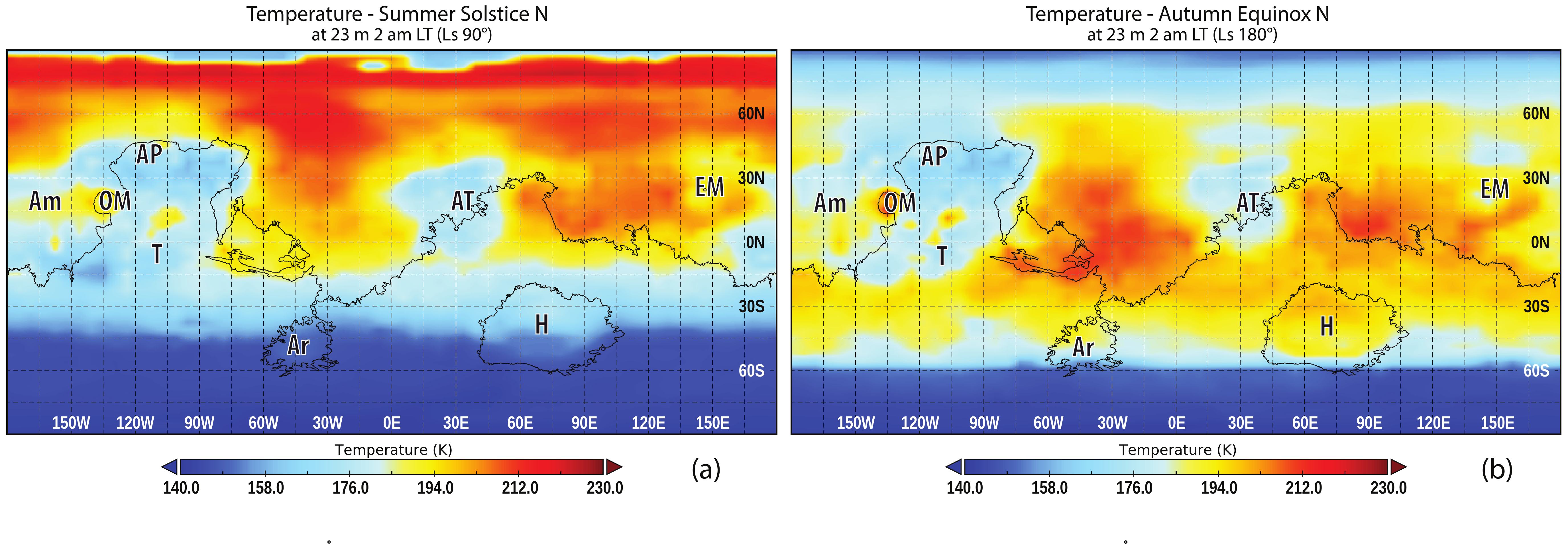}
    \caption{Seasonal dependence of nighttime temperature from GCM data at approximately 23 m above the surface at the time of northern summer solstice (a) and northern autumn equinox (b). Local time is 2 am at the whole planet. The three humid zones defined previously in Section \ref{sec:results} are still identifiable here, but the contrast between the temperatures here and their warmer environment is less significant. While the difference was roughly 50 K between the coldest part of the humid zones, east from Alba Patera (AP), and west from Tharsis (T) in Figure \ref{fig:gcm_ts}a., here at approximately 23 m high above the surface the maximum deviation is around 30 K. Compared with the surface data it is overall warmer during all seasons.} 
   \label{fig:gcm_temp23}
\end{figure}

In Figure \ref{fig:gcm_rh23m}, the relative humidity calculated from GCM values is shown, approximately 23 $m$ above the surface. We investigated the relative humidities at this height to see how much variation does a $\sim$ 20 $m$ difference cause. Looking at the results, the distribution of relative humidity can be quite different close to the surface, as expected. The global ``trend" is somewhat similar to what we have seen in the near surface relative humidity figures (Figure \ref{fig:gcm_rh_2am}), the southern summer is drier than the northern summer, in agreement with \citep{tokano2005,smith2002,jakosky1982}. However, while there are slightly higher relative humidity values at the presumed ``humid" regions during the northern summer, there is barely any visible variation in the northern winter solstice figure concerning these regions. In Figure \ref{fig:gcm_temp23} we can see the temperature values modelled with GCM at approximately 23 m above the surface 2 am LT, at the time of northern summer (a) and northern autumn (b). In these temperature maps the three humid zones defined in section \ref{sec:results} are still visible, but the differences between these regions and their surroundings are less notable. Comparing these figures with the aforementioned Figure \ref{fig:gcm_rh23m}, the substantial role of near surface effects in the behaviour of relative humidity at 4 m height is apparent. \\

\begin{figure}[H]
    \centering
    \includegraphics[width=\columnwidth]{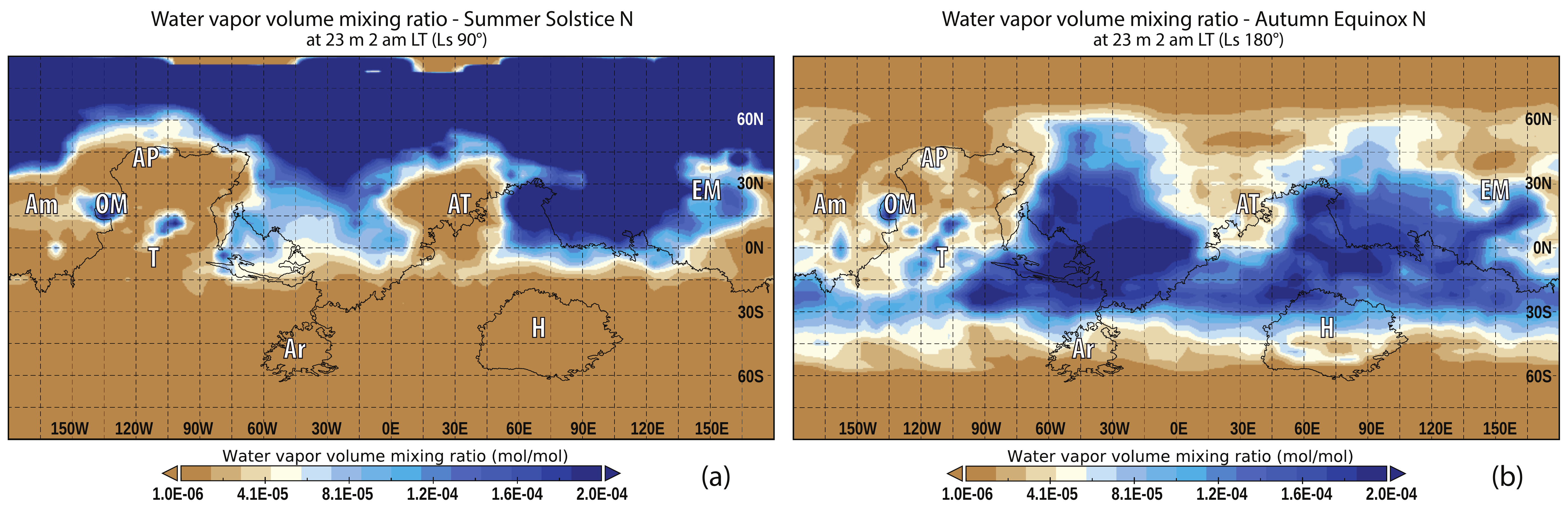}
    \caption{Seasonal dependence of water vapor volume mixing ratio at approximately 23 m above the surface from GCM model calculations. The local time is 2 am globally at northern summer solstice (a) and northern autumn equinox (b). The three humid zones defined in Section \ref{sec:results} are visible here as well. Comparing these figures with the ones 4 m above the surface (Figure \ref{fig:gcm_vmr}) there are no considerable differences. } 
    \label{fig:gcm_vmr23}
\end{figure}

In Figure \ref{fig:gcm_vmr23} the water vapor volume mixing ratio is shown at approximately 23 m above the surface. In inset (a) we can see a 2 am snapshot at the time of the northern summer solstice, and in inset (b) at the time of the northern autumn equinox. The three humid zones defined in Section \ref{sec:results} show up here as areas, where the VMR levels are lower than in their environment, with the exception of Olympus Mons (OM), Tharsis Montes (northwest from Tharsis denoted by T) and Elysium Mons (EM). Comparing the resulting VMR figures at 23 m high with the values modelled at 4 m above the surface (Figure \ref{fig:gcm_vmr}), there is no significant variation. The difference in values between the temperature, pressure or water vapor volume mixing ratio at the surface and at 23 m above does not explain the magnitude of variation between 4 m (Figure \ref{fig:gcm_rh_2am}) and 23 m (Figure \ref{fig:gcm_rh23m}) in the global relative humidity maps. Thus we conclude that the thermal inertia likely plays a great role in determining near-surface relative humidity levels. \\

\section{Discussion}
\label{sec:discuss}

We have found three areas on Mars where the relative humidity levels are elevated, potentially supersaturated regardless of season. Our findings of high relative humidity, low water vapor bearing areas coinciding with regions showing low thermal inertia and high albedo values is in good agreement with previous research in Martian near-surface and subsurface water content studies: \citet{rivera2018} shows, that the shallow subsurface for terrains with low thermal inertia ($\Gamma \lesssim 300 J m^{-2} K^{-1} s^{-1/2}$) may be occasionally favorable to brine formation through deliquescence, and the localization of elevated relative humidity regions agree quite well when comparing the maps of seasonally averaged water vapor column abundance maps shown in \citep{smith2002,tokano2003}. These ``humid zones" cool down strongly, which leads to the possible oversaturation making them interesting in relation to deliquescence processes. Low thermal inertia values, beside causing strong cooling and oversaturation, also suggest very fine soil particles with high specific surface area, which may enhance the adsorbing ability of these regions due to a larger adsorptive capacity \citep{tokano2003}. The ``humid zones" we identified are also in good agreement with the potential solubility of O$_2$ in brines (especially containing perchlorates) across the Martian near-surface \citep{stamenkovic2018}. The Tharsis region and Arabia Terra have been also previously identified as possibly containing higher amounts of adsorbed water \citep{tokano2003}. Considering the low nightly surface temperatures shown in Figure \ref{fig:gcm_ts} and the drop in the water vapor content above the surface (Figure \ref{fig:vmrd}), during the night water vapor likely condensates on the surface \citep{moores2011,whiteway2009}. \\

Nowadays on Mars the largest seasonal source of atmospheric water comes from the sublimation of the H$_2$O polar caps in summer \citep{richardson2002b,bass2000}. The sublimation of the northern polar cap begins around late northern spring, and continues throughout northern summer. During this season the sublimed water vapor gradually gets transferred from the polar regions towards the equator. Towards the end of the season, water vapor gets transferred from the Southern Hemisphere northwards, and in the Northern Hemisphere southwards because of the changing mean meridional circulation, however the net transport is still southward \citep{steele2014}. This process can be seen in the relative humidity maps (Figures \ref{fig:gcm_rh_2am} and \ref{fig:tes_rh}) with the relative humidity levels rising as we move towards northern autumn. It is interesting to note, that the main transport routes are Arabia Terra and Tharsis  \citep{steele2014}, which regions are also parts of the three ``humid" zones defined previously. As we reach the northern autumn equinox, in the relative humidity figures we can observe a band at 50$\degree$ S latitude, where it is about an order of magnitude more humid, than elsewhere in the vicinity. There is also another humid band at 60$\degree$ S latitude, where the relative humidity levels are even higher, probably associated with transient eddies and the increased water vapor supply as the seasonal ice deposits around the southern pole also begin to sublime. A similar effect can be seen at 60$\degree$ N latitude during northern spring and summer. The south pole reaches its maximum water vapor column value around the time of the northern winter solstice, Ls 270$\degree$. There is some elevation in relative humidity levels during this time in the Southern Hemisphere, but compared to the other seasons, the regions south from 15$\degree$ S are dry. This could be the consequence of the intensifying Southern Hemisphere Hadley cell, which leads to transport from the southern polar area towards the winter hemisphere being limited. This together with the lower temperatures cause water vapor to recondense back onto the southern pole. As can be seen in the relative humidity figures, by the time of the northern spring equinox the majority of the water vapor is again mostly restricted to the mid-latitudes, with transient eddies being responsible for transport polewards from 45$\degree$ \citep{steele2014}. \\

Our work concentrated on the global distribution of elevated relative humidity values and their seasonal behaviour. Because of the complexity of modelling the surface-atmosphere interactions, the water vapor volume mixing ratio is currently reliable only until approximately 4 m above the surface. Assuming the water vapor to be well mixed between the surface and 4 m is a good estimation, but due to comparing atmospheric vapor values with surface temperature data, the numerical values can reach unrealistically high saturation levels in some cases. The nighttime atmosphere is much warmer than the surface, especially above regions covered with low thermal inertia bearing materials, and just near it the atmosphere is drier due to the interaction with the surface. In some way our work shows a ``potential supersaturation" and while the specific numerical values may not be accurate in all instances, the distribution of elevated relative humidity is reliable and gives a good estimate even with its known limitations. This study could be useful to study the global behaviour of near-surface relative humidities and the role of thermal inertia in it, and also to provide a base in choosing future sites of interest in relation to e.g. deliquescence processes. At the identified ``humid" zones the surface condensation of H$_2$O is enhanced, thus nighttime frost formation is more likely than at other locations at the same latitudes. The process of deliquescence is also more probable at these regions, thus their further analysis related to the potential of liquid water or brine formation and the occurrence of hyroscopic minerals there is important. \\

\section{Conclusion}
\label{sec:conc}

We investigated the global seasonal and daily variations of near-surface relative humidity on Mars using calculations from modelled and measured data. We mostly focused on snapshots during the night, particularly approximately 2 am everywhere on the planet. The reason behind this was to eliminate daily effects related to differences in insolation, and to be able to compare relative humidity derived from LMDZ GCM model calculations with values calculated from TES measurements. We mainly focused on global effects, but also touched upon the subject of daily variations of relative humidity and water vapor volume mixing ratio for illustration purposes. \\

We identified three ``humid" zones, which show elevated relative humidity levels compared to other areas during nighttime regardless of season, with certain variations discussed in detail in the results section. The term ``humid" refers to the relative humidity levels being generally 2-3 orders of magnitude higher than in their vicinity around 2 am local time, although it does not necessarily mean that there is a large mass of water in the atmosphere. These three areas are most likely regions on Mars, where the thermal inertia is low and the albedo is high, which suggest that they are covered by unconsolidated fine dust with grain sizes less than $\sim$ 40 $\mu$m. As a result they show larger temperature variations throughout a day, cool down more quickly during the night and heat up more rapidly during the day. This effect can be seen on the 2 am LT maps, where the ``humid" zones show 30-50 K lower temperatures than their surroundings. Above these areas the water vapor volume mixing ratio drops quickly during the night with 1-2 orders of magnitude. The difference in values either between the surface temperature and $\sim$ 23 m high temperature or the $\sim$ 4 m high and $\sim$ 23 m water vapor volume mixing ratio does not explain the magnitude of variation seen comparing the near-surface relative humidity and $\sim$ 23 m relative humidity levels. At $\sim$ 23 m above the surface mainly the zonal and meridional circulation patterns dominate the distribution of relative humidity values, with little variation in connection to thermal inertia of the surface. For this we conclude, that thermal inertia could play an important role in the changes of near-surface relative humidity, and in the overall humidity evolution of the atmospheric layers very close to the surface. As the water vapor levels drop near-surface during the night, we suggest that some kind of condensation of water vapor could occur during the night supported by the elevated relative humidity here in the form of brines, wettening of the fine grains or deliquescence. These findings confirm that surface TI dominate the locations of nighttime condensation below the planetary boundary layer and it causes an observable decrease in the atmospheric H$_2$O mass there. This process could influence the global daily H$_2$O exchange in the very shallow regions of the regolith and may be examined by in-situ measurements according to the locations and periods when and where the probability of H$_2$O condensation is the highest. Examining the daily relative humidity curves at the three ``humid" zones and two reference ``dry" zones, it is observable, that the relative humidity values are higher during the night and lower during the day at Alba Patera and Arabia Terra locations. Elysium Mons also shows higher nightly and lower daily RH values, but the difference to the reference areas is not that substantial. A more detailed investigation of the daily variations will be a part of our future research in near-surface relative humidity. Precisely modelling and predicting the humidity of near-surface layers of Mars is not a trivial task, but we believe that our work is reasonable and could be useful in the planning of future research in the topic of the possibility of liquid water on present day Mars. \\

\section{Acknowledgement}
\label{sec:ack}

This work was funded by the COOP-NN-116927 project of NKFIH, EXODRILTECH (4000119270) project, the GINOP-2.3.2-15-2016-00003 grant of the National Research, Development and Innovation Office (NKFIH, Hungary) and the TD1308 \textit{Origins and evolution of life on Earth and in the Universe} COST actions number 39045 and 39078. Special thanks to Vladimir V. Zakharov for his invaluable help with the NetCDF file handling. The NetCDF files were visualized with the NASA GISS Panoply viewer developed by Dr. Robert B. Schmunk. The thermal inertia map (Figure \ref{fig:tesTI}.) was created by Vilmos Steinmann. We also would like to thank the referees for their valuable input which have greatly improved this manuscript. \\

\appendix

\section{TES relative humidity calculation}
\label{app:tes}

To obtain the water vapor mixing ratio, we divided the number of H$_2$O molecules per cm$^2$ in an atmospheric column by the number of CO$_2$ molecules per column. This is an approximation, but since roughly 95\% of the martian atmosphere is composed of carbon dioxide, this assumption is reasonable. The number of CO$_2$ molecules can be calculated by dividing the mass of CO$_2$ per cm$^2$ by the mass of a single molecule: 

\begin{equation}
    n_{CO_2} = \frac{m_{CO_2}}{44 \times 1.66 \times 10^{-24}}
\end{equation}

\noindent Assuming hydrostatical equilibrium in the atmosphere, the mass of CO$_2$ in a column is equal to the pressure between two atmospheric levels divided by the gravitational constant on Mars (374 cm/s$^2$): 

\begin{equation}
    n_{CO_2} = \frac{\mathrm{p_{surf}} - \mathrm{p_{top}}}{374 \times 44 \times 1.66 \times 10^{-24}} = \left( \mathrm{p_{surf}} - \mathrm{p_{top}} \right) \times 3.6595 \times 10^{19} \, \mathrm{molecules/cm^2}
\end{equation}

\noindent We can obtain the number of H$_2$O molecules in a similar way: 

\begin{equation}
    n_{H_2O} = \frac{m_{H_2O}}{18\times 1.66 \times 10^{-24}}
\end{equation}

\noindent The water vapor is given in precipitable microns in the TES measurement data, which is the liquid water equivalent in microns. The density of liquid water is 1.0 g/cm$^3$ and a micron equals $1\times10^{-4}$ cm, so the number of water molecules is given by: 

\begin{equation}
    n_{H_2O} = \frac{H_2O_{\mathrm{column}}\times1\times10^{-4}}{18\times1.66\times10^{-24}} = H_2O_{\mathrm{column}} \times 3.3456 \times 10^{18} \mathrm{molecules/cm^2}
\end{equation}

\noindent To get the water vapor volume mixing ratio, we divide the two values: 

\begin{equation}
    Q_0 = \frac{n_{H_2O}}{n_{CO_2}} = \frac{H_2O_{\mathrm{column}}\times 3.3456 \times 10^{18}}{\left( \mathrm{p_{surf}} - \mathrm{p_{top}} \right) \times 3.6595 \times 10^{22}} = \frac{9.142 \times 10^{-5} \times H_2O_{\mathrm{column}}}{\mathrm{p_{surf}} - \mathrm{p_{top}}}
\end{equation}

%% References


%\bibliographystyle{elsarticle-num}
 \bibliographystyle{elsarticle-harv}
\section*{References}
\bibliography{ref}

\end{document}